%% file: main.tex
\documentclass[conference,a4paper,onecolumn,12pt,twoside]{IEEEtran}
\usepackage{array,booktabs,amsmath,graphicx,fancyvrb,tabularx, longtable}
\usepackage[tt=false, type1=true]{libertine}
\usepackage[varqu]{zi4}
\usepackage[libertine]{newtxmath}
\usepackage[tableposition=top]{caption}
\usepackage[utf8]{inputenc}
\usepackage{svg}
\usepackage{wrapfig}
\usepackage{subcaption}
\usepackage{booktabs}
\usepackage{hyperref}
\usepackage[noabbrev]{cleveref}
\usepackage[top=1.22in, bottom=1.15in, left=0.75in, right=0.75in]{geometry}
\usepackage{url}
\usepackage[numbers]{natbib}
\usepackage{calc}  
\usepackage{enumitem}  
\usepackage{tabulary}
\usepackage{caption}
\newcommand{\rot}[1]{\rotatebox{90}{\parbox{2cm}{\tiny \hyphenpenalty=100000 \raggedright #1}}}

\usepackage{verbatim}
\usepackage{subcaption}
\usepackage{adjustbox}
\DeclareMathOperator*{\argmax}{arg\,max}

\usepackage{tabulary}

\title{A probabilistic database approach to autoencoder-based data cleaning}

\author{\IEEEauthorblockN{R.R. Mauritz}
\IEEEauthorblockA{\textit{University of Twente}\\
\textit{Faculty of EEMCS} \\
Enschede, NL \\
r.r.mauritz@student.utwente.nl}
\and
\IEEEauthorblockN{F.P.J. Nijweide}
\IEEEauthorblockA{\textit{University of Cambridge}\\
\textit{Computer Lab} \\
Cambridge, UK \\
fpjn2@cam.ac.uk}
\and
\IEEEauthorblockN{J. Goseling}
\IEEEauthorblockA{\textit{University of Twente}\\
\textit{Faculty of EEMCS} \\
Enschede, NL \\
j.goseling@utwente.nl}
\and
\IEEEauthorblockN{M. van Keulen}
\IEEEauthorblockA{\textit{University of Twente}\\
\textit{Faculty of EEMCS} \\
Enschede, NL \\
m.vankeulen@utwente.nl}
}

\crefname{paragraph}{section}{sections}%

\usepackage{fancyhdr}
\usepackage{etoolbox}
\usepackage{changepage}

\providecommand{\keywords}[1]
{{
  \small	
  \textbf{\textit{Keywords---}} #1
}}

\usepackage{fullwidth}

\begin{document}
\begin{fullwidth}[innermargin=-0.75in,width=\linewidth+1.5in,nobreak]

\maketitle

\end{fullwidth}

\begin{abstract}
Data quality problems are a large threat in data science. 
In this paper, we propose a data-cleaning autoencoder capable of near-automatic data quality improvement. It learns the structure and dependencies in the data and uses it as \emph{evidence} to identify and correct doubtful values. We apply a probabilistic database approach to represent weak and strong evidence for attribute value repairs. A theoretical framework is provided, and experiments show that it can remove significant amounts of noise (i.e., data quality problems) from categorical and numeric probabilistic data. Our method does not require clean data. We do, however, show that manually cleaning a small fraction of the data significantly improves performance.
\end{abstract}

\hspace{10pt}
\keywords{data cleaning, probabilistic databases, autoencoders}

\input{introduction} 
\input{preliminaries}
\input{problem}
\input{methodology}
\input{experiments}

\input{conclusions}

\bibliography{Bibliography} 
\bibliographystyle{ieeetr}

\end{document}

%% file: introduction.tex
\section{Introduction}\label{sect:introduction}
Data quality problems are a large threat in data science \cite{halevydataspaces}. Many techniques and methods have been proposed for detecting, managing, and resolving data quality problems (see, e.g., \cite{dqhandbook,diqbattini,Ilyas2019,synthesiscleaning}) as well as for being able to meaningfully use data that still contains unavoidable problems \cite{roadtowards}. Data cleaning is a multi-faceted problem, because there are many different kinds of data quality problems and causes for these problems. Specifically, when data is the result of measurement, missing values and noise can be expected. Also, in the field of \emph{data-integration}, i.e., combining several data sources into a single and unified view \cite{lenzerini2002}, often inconsistencies and ambiguities are encountered when extracting, combining, and merging data.
In this paper, we focus on cleaning missing and incorrect categorical and numerical attribute values.

The main intuition behind our approach is the observation that data is typically `generated' by real-world processes causing patterns to exist in the data. These patterns can be used as evidence for doubting certain values (i.e., detecting a possible data quality problem) and for finding a most likely value. In the database area, given or detected (conditional) functional dependencies and other constraints have often been used for data cleaning \cite{cfds,holoclean}. The concept of a \emph{repair} for the purpose of Consistent Query Answering (CQA) is similarly based on respecting consistency constraints~\cite{synthesisrepair,wijsencqarepairs,guideddatarepair}.

Machine learning (ML) is very good at finding patterns in data, hence there is increasing interest in using ML for data cleaning with the aim to improve the quality of the cleaning result as well as to improve scalability \cite{scaredbertiequille}. In this paper, we focus on autoencoders, which are traditionally used for dimensionality reduction and denoising~\cite{Goodfellow-et-al-2016}, because of their ability to find patterns in an unsupervised manner or with only a limited amount of ground truth data.

Common to most approaches for data cleaning, including ML-based ones, is that there is an underlying optimization problem targeted at finding the `most likely' data among many possibilities. However, as also observed by \cite{holoclean}, the most likely data is not necessarily the correct data. Moreover, information about these possibilities is considered an important result of a cleaning or integration process \citep{magnanisurvey-jdiq2010}. For this purpose, probabilistic database (PDB) techniques are increasingly applied to be able to represent possible alternatives with their likelihoods \cite{roadtowards,vankeulen2018,holoclean,dqhandbook,panseindeterministic}. In this paper, we apply PDB techniques for representing weak and strong recommendations for attribute value repairs. We propose a method whose output can be used directly when stored in a PDB, or it can be used in a subsequent decision step for deciding which alternative values to pick or what values to manually inspect. Our method accepts as input both probabilistic data as well as ordinary `crisp' data. Closest to our work are HoloClean \cite{holoclean} and SCARE \cite{scaredbertiequille}. HoloClean uses a variety of signals from given constraints and the repairs that are possible for obtaining a probabilistic model capturing the uncertainty. Similar to SCARE, we do not rely on given constraints, but assume their effect is present as patterns in the data. HoloClean's probabilistic inference method produces marginal probabilities for possible repairs similar to our proposed method. Our approach relies on a different kind of probabilistic model and inference than HoloClean and SCARE, namely autoencoder-based.

\paragraph{Contributions} 

\begin{itemize}
\item A \emph{data-cleaning autoencoder (DCAE)} approach for cleaning categorical and numeric attributes. The basic approach is unsupervised, meaning that it does not require clean data (i.e. correct without uncertainty) for training.
\item An extension of the basic approach in which a small fraction of the records is manually cleaned. This semi-supervised approach uses a limited amount of clean records as labelled ground truth data. 
\item A method for generating synthetic test data with embedded dependencies for the purpose of evaluating our cleaning approach.
\item Experimental evaluation of the cleaning performance under many varying circumstances: levels and kinds of uncertainty and errors, size of the database, and several data parameters and architecture hyper-parameters.
\item Experimental results on real-world datasets. 
\end{itemize}

\paragraph{Outlook}
\Cref{sec:preliminaries} provides related work on probabilistic databases and autoencoders, introduces notation, and describes the intuition behind applying autoencoders for the purpose of data cleaning. \Cref{problem-formulation} describes the proposed solution by formalizing its core as a machine learning problem. \Cref{sec:methodology} introduces our evaluation framework and experimental setup. \Cref{sec:experiments} presents the experimental results and we conclude with \Cref{sec:conclusions}, containing conclusions and future work.

%% file: preliminaries.tex
\section{Preliminaries and related work}
\label{sec:preliminaries}

\subsection{Probabilistic databases}\label{sec:pdbs}

Various probabilistic databases have been proposed. Examples include MayBMS
\citep{antova2009maybms}, Trio \citep{widom2004trio}, and
MCDB \citep{jampani2008mcdb} for uncertain relational data, and IMPrECISE \cite{deKeijzer2008,keulenvldbj09} and others \cite{exprpxmlvldbj09} for uncertain XML. Also, probabilistic logics have been defined to capture and reason with uncertain information \citep{fuhr2000probabilistic,
judged2016, DeRaedt2015}. There is much variety in how the uncertainty is modelled in these systems. For example, MayBMS's U-relations \cite{maybmsurelations} focus on tuple-level
uncertainty where probabilities are attached to tuples, while MCDB focuses
on attribute-level uncertainty where a probabilistic value generator
function captures the possible values for the attribute. These uncertainty models vary in expressiveness \cite{exprpxmlvldbj09}. In models that only attach probabilities to tuples, the uncertainty of the tuples is inherently independent of each other. In contrast, the world set descriptors of MayBMS and the descriptive sentences of JudgeD \cite{judged2016} also allow them to express complex dependencies involving full dependence and mutual exclusion \cite{revisiting2015} necessary for faithfully capturing the possible outcomes of a data integration process.

\subsection{Uncertainty model and assumptions}\label{sect.pdb_modeling}
We start with presenting a model for categorical (nominal) data. 
Uncertainty of an attribute value is modelled as a probability distribution over all possible values for that attribute, i.e., with a probability for each possible value summing up to 1. To illustrate, the uncertainty that the colour of a car is green or blue, but not red, is modelled as assigning probabilities 0.5, 0.5, and 0 to the possible values ``green'', ``blue'', and ``red'', respectively.

We use the following notation for our uncertainty model:
\begin{itemize}
    \item Without loss of generality, we restrict our attention to a single table with $N$ attributes and $M$ records.
    \item Let $\mathcal{A}$, $|\mathcal{A}|=N$, be the set of names of the attributes. If not stated otherwise, we assume $\mathcal{A} = \{1, 2, \dots, N\}$.
    \item All attributes are categorical. Attribute $j\in\mathcal{A}$ takes values in $\mathcal{K}_j$. Let $K_j=|\mathcal{K}_j|$, i.e., attribute $j$ can take $K_j$ different  categories. If not stated otherwise, we assume $\mathcal{K}_j = \{1, 2, \dots, K_j\}$. 
    \item The likelihood of the value of attribute $j$ in record $i$ is represented as a probability mass function $p_{ij}$ over $\mathcal{K}_j$, where $p_{ij}(k)$ is the probability that the attribute has value $k\in\mathcal{K}_j$.
    \item Let $x_i$ denote the $i$-th record. In the remainder it will be useful to represent $x_i$ as the concatenation of all probability mass functions, i.e., if $\mathcal{A}=\{1, 2, \dots, N\}$ and $\mathcal{K}_j = \{1, 2, \dots, K_j\}$ for all $j\in\mathcal{A}$ we obtain
    \begin{equation}\label{eq.record}
        x_i
            = \left(p_{i1}(1), p_{i1}(2), \dots, p_{i1}(K_1), p_{i2}(1),\dots,p_{i2}(K_2), p_{i3}(1), \dots, p_{iN}(K_N)\right).
    \end{equation}
\end{itemize}

\begin{table}[t]\centering\footnotesize \begin{tabular}{|l|l|l|l|l|}
\hline
           & \multicolumn{2}{l|}{\textbf{Eye colour}} & \multicolumn{2}{l|}{\textbf{Hair colour}} \\ \hline
           & \textit{Blue}      & \textit{Brown}      & \textit{Light}       & \textit{Dark}      \\ \hline
\textbf{1} & 0.7                & 0.3                 & 1.0                  & 0.0                \\ \hline
\textbf{2} & 0.8                & 0.2                 & 0.9                  & 0.1                \\ \hline
\textbf{3} & 0.0                & 1.0                 & 0.5                  & 0.5                \\ \hline
\end{tabular}
\caption{Example Probabilistic database (PDB).}
\label{tab:example_db}
\end{table}

We illustrate our notation through the example database that is depicted in \Cref{tab:example_db}.
In this database we have:
\begin{equation}
    \mathcal{A} = \{\text{eye colour}, \text{hair colour}\},
\end{equation}
\begin{equation}
    \mathcal{K}_{\text{eye colour}} = \{\text{blue}, \text{brown}\},
    \quad
    \mathcal{K}_{\text{hair colour}} = \{\text{light}, \text{dark}\},
\end{equation}
and, for instance
\begin{gather}
    x_1 = (0.7, 0.3, 1.0, 0.0),\\
   p_{1, \text{eye colour}} = (0.7, 0.3), \\
   p_{1, \text{eye colour}}(\text{blue}) = 0.7.
\end{gather}

The above model covers only categorical attributes. For probabilistic databases, extensions to the discrete categorical distribution model exist, allowing for the use of probabilities with a continuous distribution~\cite{Grohe2019}. In \Cref{sec:categorization} we explain how we deal with continuous attributes.

\subsection{Autoencoders}\label{ssec:autoencoders}
The \textit{autoencoder} (AE) dealt with in this paper is a feedforward, non-recurrent neural network having an input layer, several hidden layers and an output layer with the same number of nodes as the input layer.
An AE is meant to learn the structure and patterns in the input data to reproduce its input according to this structure and these patterns.
To achieve this, constraints (such as a reduced dimensionality of the middle layer) are added to force the network to learn a representation of the training data with a reduced feature space.
An AE is usually an \emph{unsupervised} learning method, as no other prior knowledge about the data (i.e., in terms of targets) is required for this process \cite{Vincent2010}.
A typical use of an AE is for noise-cancelling in images: given an input image, it can learn to reproduce the image in the output without the noise. This can be achieved using only unsupervised learning, but performance can be improved by using semi-supervised or supervised learning, where the intended output of the AE is used during training.

An AE consists of an encoder $g_{\phi}(\cdot): \mathcal{X}\to\mathcal{H}$ and decoder $f_{\theta}(\cdot):\mathcal{H}\to\mathcal{X}$, parameters $\phi$ and $\theta$ are the weights and biases of the encoder and decoder, respectively. These parameters are trained via minimizing the loss function $\mathcal{L}$ which is a measure for reconstruction error: 
\begin{equation}\label{eq.ae_optim}
    \phi, \theta = \arg\min_{\phi,\theta}\mathcal{L}\big(Y,f_\theta(g_\phi(X))\big),
\end{equation}

where $Y$ represents the desired output, $X$ represents the set of training data and $f_{\theta}(g_{\phi}(X))$ represents the output to the AE based on input $X$. In the case of unsupervised learning (the standard use case for AE's), the desired output is taken as $Y = X$. In a supervised setting, for instance, in image denoising $X$ represents noisy image and the corresponding noise-free version.

As mentioned, an AE is well known for its dimensionality reduction and noise-cancelling ability. Other uses for AEs include anomaly (outlier) detection, where the input is determined to be an anomaly if the network is unable to reconstruct the input \cite{Sakurada2014}. Note that nothing in \Cref{eq.ae_optim} prevents the AE from not learning the identity function, which is an extreme case of overfitting. Several types of AEs exist with designs that mitigate this problem. These types of AEs are not mutually exclusive, and they may (and in fact, we do so in our DCAE) be combined \cite{Fan2017}:
\begin{itemize}
\item  An \emph{undercomplete autoencoder} has a hidden middle layer with a lower dimensionality than the input or output spaces. This implicit regularization ensures that the AE has to learn to capture the most important features from the data for it to be able to reconstruct its input well.  
\item A \emph{denoising autoencoder} has noise added to the input data before being fed to the network. The AE then is forced to learn how to remove this noise because the loss function compares the output with the original, clean input data. This prevents overfitting and enhances its noise-cancelling capabilities \cite{Vincent2010}. \item A \textit{sparse autoencoder} adds a sparsity penalty to the training criterion. The AE is now also penalized on the number of active neurons in the code (middle) layer. This constraint encourages the AE to retain a more meaningful representation of the data in the code layer.
\item A \textit{variational autoencoder} (VAE) consists of an encoder section that passes both a tensor of means and a tensor of standard deviations to the decoder section,  instead of deterministic variables as is usually the case. A term consisting of the Kullback-Leibler divergence \cite{kldiv} between this distribution, and a standard normal distribution is added to the loss function. A sample is taken from this distribution and fed to the decoder section \cite{Doersch2016}. The VAE ensures that the learned latent space representation is continuous (meaning that neighbouring data points should lead to similar outputs) and complete (all inputs should lead to a sensible output). Moreover, the attributes in the latent space are orthogonal, a property that is known as disentanglement.
\end{itemize}

\subsection{Application of AEs to data cleaning}
\label{sec:application-AE-to-DQ}

Our application of AEs is data cleaning. We argue that data cleaning can be seen as \emph{noise-cancelling in records}. Whereas in images, noise is formed by unstructured deviations in the colour of pixels, data quality problems in relational data can be regarded as unstructured deviations in the values of attributes. The AE in our approach is meant to learn the structure and patterns in the relational data for the purpose of reproducing its records, suggesting adjustments to its attribute values for them to be more in line with the structure and patterns in the data set. This also motivates our use of probabilistic data as it allows the AE to indicate with probabilities weak and strong recommendations for adjustments to the attribute values.

Our DCAE can be used in various scenarios. Given a data set with suspected data quality problems, an AE can be used to identify suspicious attribute values for manual inspection. Note that it is in principle an unsupervised approach, so no laborious labelling is necessary beforehand, although we have extended our approach such that it can exploit labelled data. We show that labelling a small fraction of the significantly improves the performance. 

The approach can both be used taking `normal' data as well as taking probabilistic data as input.
In the former case, it is important to understand that normal `crisp' data is just a special case of probabilistic data, namely where one of the possible values is assigned all the probability mass in the probability distribution. In the latter case, it should be remarked that the AE in our approach abstracts from the dependencies stored in a probabilistic database (see \Cref{sect.pdb_modeling}). It takes as input probabilities derived from the world set descriptors \cite{maybmsurelations} or descriptive sentences \cite{revisiting2015} and produces `new' probabilities. It may seem that the dependencies necessary for expressing things like mutual exclusion are lost in the process. However, the dependencies can be retained by regarding the output of the AE as \emph{soft evidence} with which the probabilistic database is conditioned \cite{sum-conditioning}. For the output of the AE, a soft rule can be constructed with a trust level of $\alpha$, and then we incorporate this evidence by conditioning the probabilistic database. In a sense, the original data is trusted with a level of $1-\alpha$.

Note that the overall effect of training an AE on only noisy data, is that the structure of the data as learned by the AE includes the uncertainty and errors of this data. As long as the amount of noise is limited and unstructured (see Section \ref{ssec:addingnoise}), the AE will suggest corrections on how to remove that uncertainty from records affected by noise. Also, there may be records for which the PDI process indicated no uncertainty, but that are wrong. The corrections on these records will perturb them towards the general structure that has been learned. Effectively, the AE introduces uncertainty on these records, indicating a doubt on the correctness of these records. Finally, because all records will be perturbed towards the general structure, correct records without uncertainty will also be affected by introducing small amounts of uncertainty. This effect is expected to be limited; we verify this through numerical experiments in \Cref{sec:experiments}.

%% file: problem.tex
\section{Problem Formulation \& Proposed Solution}\label{problem-formulation}
\begin{figure} 
    \centering
    \includegraphics[width=1.0\textwidth]{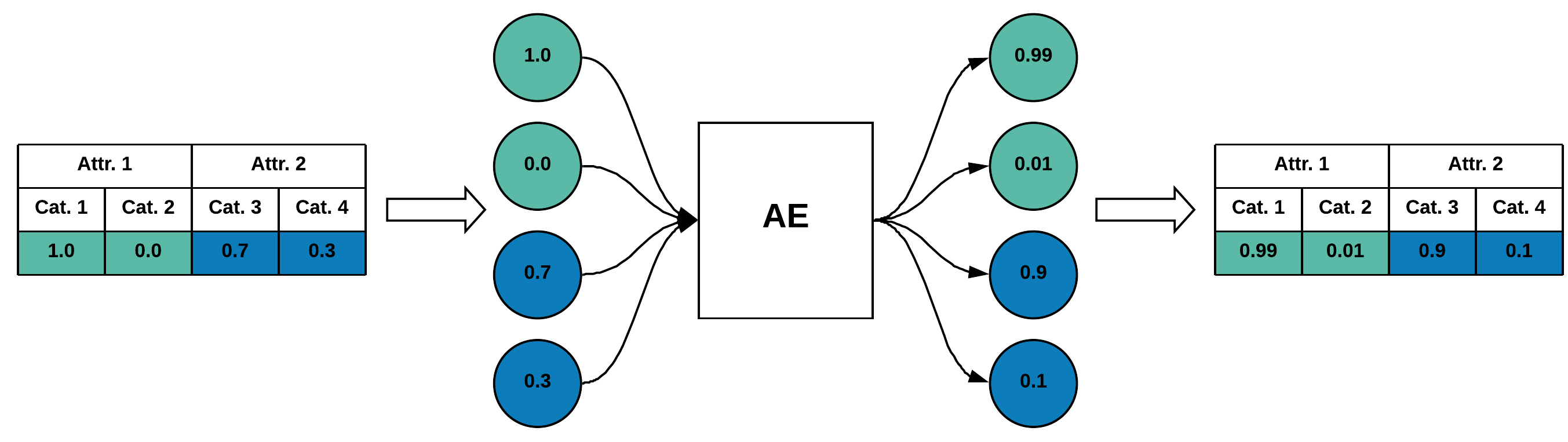}
    \caption{DCAE: Input and output representation.} 
    \label{fig:ae_sol}
\end{figure}

As explained in \Cref{sect:introduction}, our goal is to improve the data quality in a PDB by means of an AE that learns the structure of the data and is able to identify and correct outliers. We refer to such an autoencoder as a data-cleaning autoencoder (DCAE). Our idea is to use records of the probabilistic database $\mathcal{D}_{\text{PDB}}$ as input to the DCAE so that the DCAE learns the structure of this probabilistic data. The DCAE operates on a per-record basis, providing a cleaned record at its output. This is illustrated in \Cref{fig:ae_sol}. More details are provided in the remainder of this section.

\subsection{Input and output representation}\label{ssec:input_output} 
Vectors $x_i$ of the form of \Cref{eq.record} are used as input for our AE. As a result, each attribute $j\in\mathcal{A}$ has $K_j$ corresponding nodes in the input layer of the model, one for each possible category from $\mathcal{K}_j$. In total, the model then has $\sum_{j=1}^{N}K_j$ of input nodes.

We denote by $y_i$ the output corresponding to input $x_i$. The number of output nodes (i.e., dimension of $y_i$) is equal to the number of input nodes (i.e., dimension of $x_i$). Similar to \Cref{eq.record} we denote $y_i$ as
\begin{equation}
y_i
= \left(q_{i1}(1), q_{i1}(2), \dots, q_{i1}(K_1), q_{i2}(1),\dots,q_{i2}(K_2), q_{i3}(1), \dots, q_{iN}(K_N)\right),
\end{equation}
where $q_{ij}$ is the output probability distribution for attribute $j\in\mathcal{A}$, i.e., $q_{ij}$ is the cleaned version of $p_{ij}$.

\begin{figure} 
    \centering
    \captionsetup{justification=centering}
    \vspace{10px}
    \fbox{\includegraphics[width=0.6\textwidth]{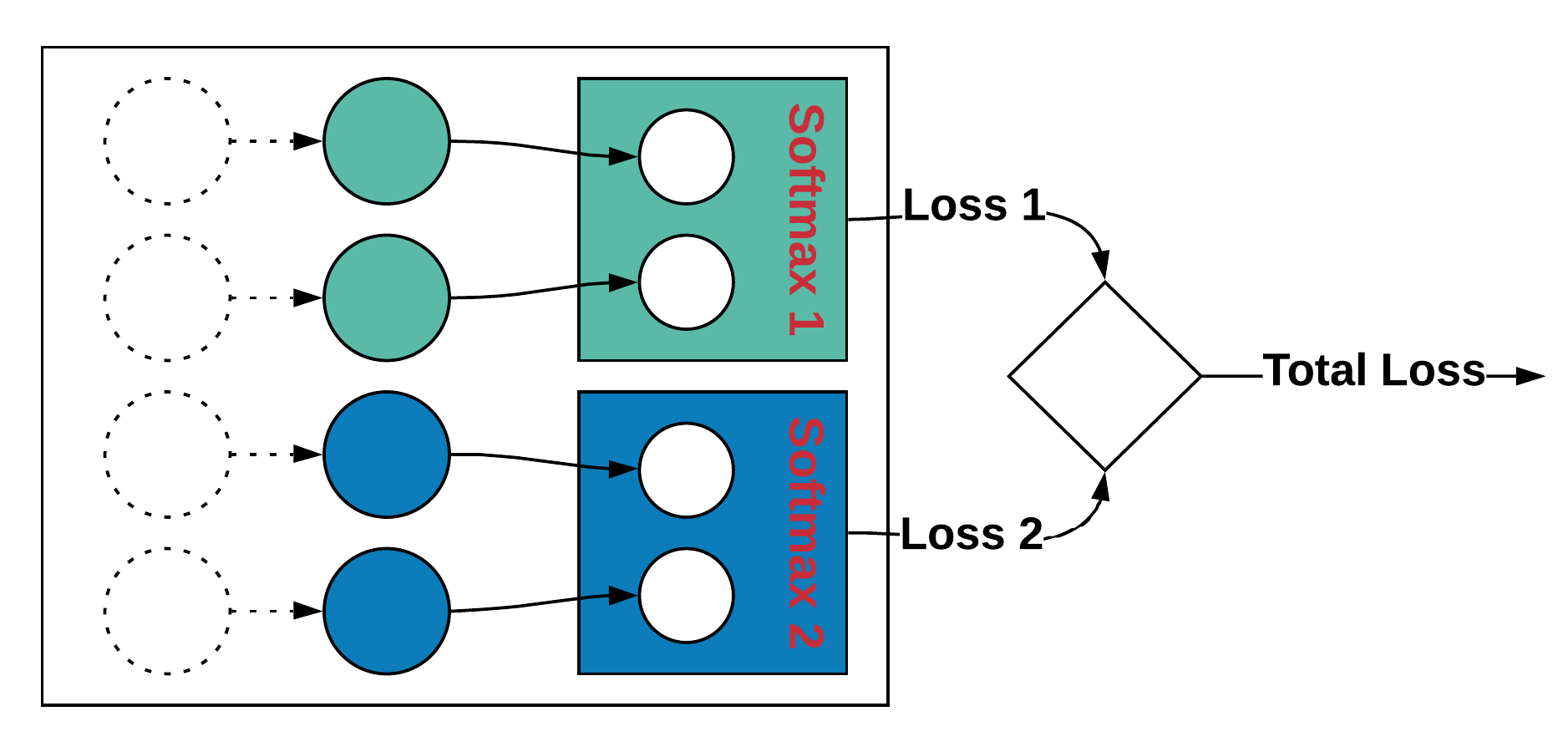}}
    \caption{Per-attribute softmax function in the output layer and overall loss function.}
    
    \label{fig:output_AE_Softmax}
\end{figure}

In order to ensure that $q_{ij}$ is a probability distribution over $\mathcal{K}_j$, the last layer in our network is a per-attribute softmax~\cite[Section~6.2.2.3]{Goodfellow-et-al-2016}, as illustrated in \Cref{fig:output_AE_Softmax}. More specifically, we let
\begin{equation}
    q_{ij}(k) = \frac{e^{\bar q_{ij}(k)}}{\sum_{k=1}^{K_j}e^{\bar q_{ij}(k)}},\quad \forall k\in\mathcal{K}_j, j\in\mathcal{A},
\end{equation}
where
\begin{equation}
    \left(\bar q_{i1}(1), \bar q_{i1}(2), \dots, \bar q_{i1}(K_1), \bar q_{i2}(1),\dots, \bar q_{i2}(K_2), \bar q_{i3}(1), \dots, \bar q_{iN}(K_N)\right)
\end{equation}
is the input to this softmax layer.

\subsection{Autoencoder architecture}\label{ssec:ae_architecture}
\begin{figure}
    \centering
    \includegraphics[width=0.8\textwidth]{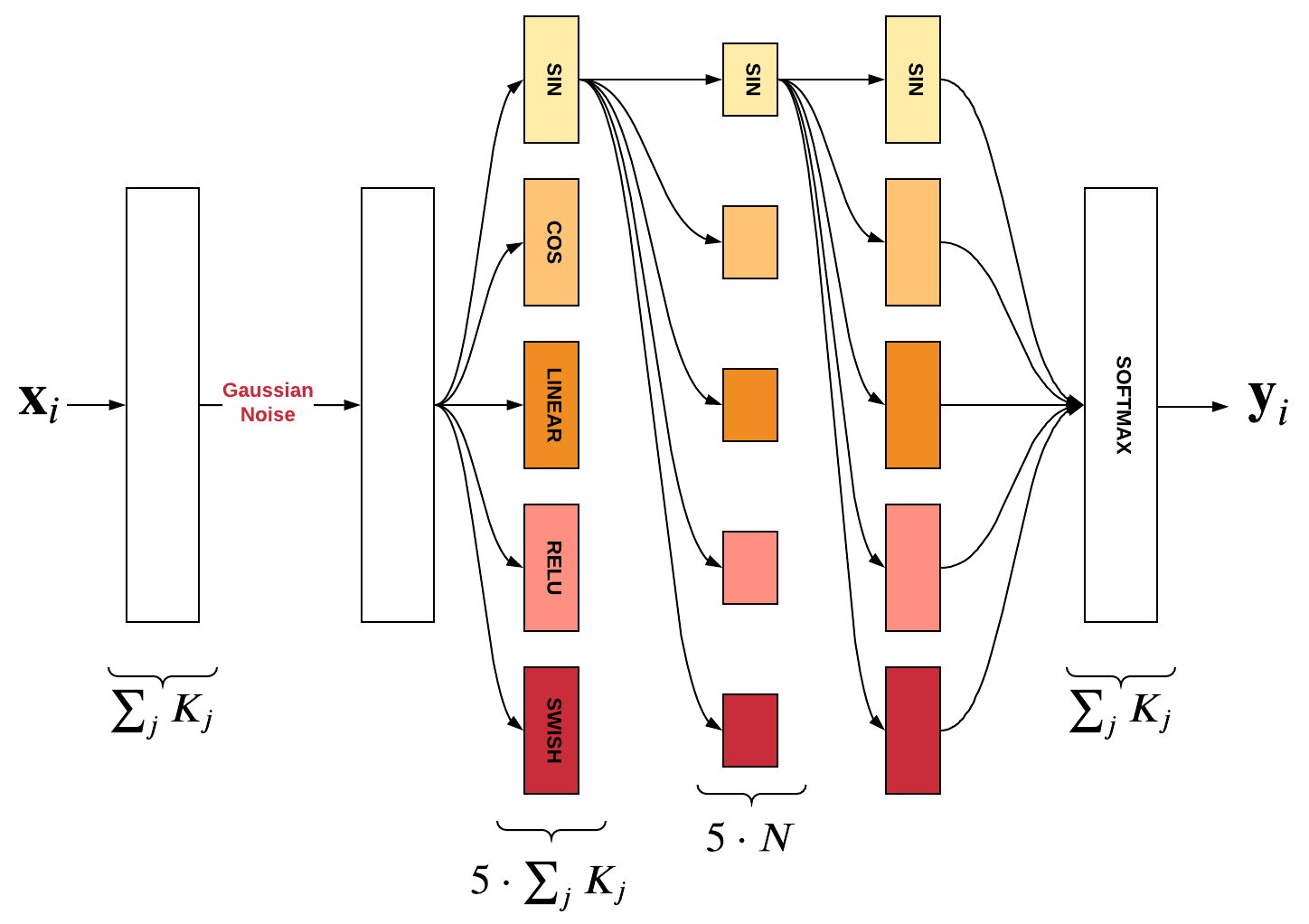}
    \caption{Default architecture of the DCAE. Gaussian noise is only added during the training process.}
    \label{fig:ae_default}
\end{figure}

The standard AE architecture that we use for our DCAE model is given in \Cref{fig:ae_default}. The AE consists of an input layer consisting of $\sum_{j=1}^{N}K_j$ nodes (see \Cref{ssec:input_output}). In (and only in) the training process, Gaussian noise is added so that the AE becomes a denoising AE that has to learn to remove the Gaussian noise, see \Cref{ssec:autoencoders}. In \Cref{sec:experiments}, we explore various parameter settings to justify our choice for the amount of Gaussian noise that we add to this layer. This layer is then fully connected to the input layers of each of five sub-channels, having the same number of nodes. Each channel uses a different and fixed activation function (sin, cos, linear, ReLU and Swish). The idea behind this is that each channel can capture a different non-linearity in the input data, which mostly linear activation functions like ReLU would not capture. This is similar to a convolutional layer leading to multiple channels capturing different structures. Then, for each sub-channel, the input layer is fully connected to the hidden layer of each sub-channel. Each hidden layer has $N$ nodes. This approach encourages the encoder section to produce one "best guess" for each of the $N$ PDB attributes as its output, like in a regression network. As a result, the decoder section is meant to learn to produce a one-hot encoding of this number. Each hidden layer from a sub-channel is again fully connected to the output layer of each sub-channel. Those five output layers are in turn fully connected to a single output layer having again $\sum_{j=1}^{N}K_j$ nodes that uses a per-attribute softmax as activation function (see \Cref{ssec:input_output}).  

We tried various modifications of this architecture in an attempt to increase performance, none of which succeeded. Turning the DCAE into a VAE (see \Cref{ssec:autoencoders}) by doubling the size of the encoder section's output to form the parameters $\mu_{\text{latent}}$  and $\sigma_{\text{latent}}$ which are used to sample from a normal $N(\mu_{\text{latent}},\sigma_{\text{latent}})$, and adding a penalty to the loss function based on this distribution's deviation from $N(0,1)$ did not lead to a reliable performance increase. Using 1D convolutional layers \cite{Goodfellow-et-al-2016} led to the best performance with only one convolutional layer, which had 64 output channels and a kernel size of 3 and a stride of 1. However, its performance was always worse than the architecture without convolutional layers. Another attempt to increase performance incorporated the use of an RBF \cite{vert2004primer} kernel to imitate a support-vector machine. The best performance using this modification was seen when using 100 landmarks, but its performance was always worse than without the RBF kernel. The details of these experiments are provided in~\cite{essay82344}.

\subsection{Loss function}\label{ssec:loss}
As the data is of probabilistic nature and more specifically consists of records that are themselves ensembles of categorical probability distributions,
it makes sense to use a loss function that can measure the distance between probability distributions.
We denote by $\mathcal{L}(x_i,y_i)$ the loss function of the DCAE, i.e.,\ the loss at input record $x_i$ and output record $y_i$. Our loss function is a summation of loss per attribute, i.e.,
\begin{equation}\label{eq:loss_per_attribute}
    \mathcal{L}(x_i,y_i) = \sum_{j\in\mathcal{A}} \mathcal{L}_j(p_{ij},q_{ij}),
\end{equation}
where $\mathcal{L}_j(p_{ij},q_{ij})$ denotes the loss for attribute $j$. This is illustrated in~\Cref{fig:output_AE_Softmax}.
The most commonly used probabilistic loss function is the categorical cross-entropy loss, often called the ``log loss''~\cite[Chapter~4.3.2]{bishop2006pattern}, defined as
$-\sum_{k\in\mathcal{K}_j} p_{ij}(k)\log q_{ij}(k).$
Intimately related, and in fact, identical up to a constant~\cite{Goodfellow-et-al-2016} to the log loss, is the Kullback–Leibler divergence\cite{kldiv}, defined as 
\begin{equation}
    D_{KL}(p_{ij}\parallel q_{ij})
    = \sum_{k\in\mathcal{K}_j} p_{ij}(k)\log\frac{p_{ij}(k)}{q_{ij}(k)}.
\end{equation}
If $q_{ij}(k)=0$ and $p_{ij}(k)\neq 0$ for some $k$, $D_{KL}(p_{ij}\parallel q_{ij})$ is defined as $\infty$. For our application in a DCAE, this is troublesome because $q_{ij}(k)=0$ corresponds to the often encountered situation of no uncertainty for attribute $j$ in record $x_i$. Having extremely large (or $\infty$) values for our loss function hampers learning. Therefore, we use the Jensen-Shannon divergence (JSD)\cite{Lin1991}, which circumvents this problem. It is defined as
\begin{equation}\label{eq:jsd_as_two_kl}
    \mathcal{L}_j(p_{ij},q_{ij}) = JSD(p_{ij}\parallel q_{ij}) = \frac{1}{2}D_{KL}(p_{ij}\parallel r_{ij}) + \frac{1}{2}D_{KL}(q_{ij}\parallel r_{ij}),
\end{equation}
where $r_{ij}(k)=\left(p_{ij}(k)+q_{ij}(k)\right)/2$.
The JSD measures how different probability distributions are; a larger JSD means a larger difference. By using this JSD as loss function, the AE learns to minimize the difference between the probability distributions $p_{ij}$ and $q_{ij}\ \forall j\in\mathcal{A}$.

When evaluating the performance of our approach we calculate the loss between two probabilistic databases $X$ and $Y$. This is defined as the sum of the loss value over the individual records, i.e.,
\begin{equation}\label{eq:loss_per_PDB}
    \mathcal{L}(X,Y) = \sum_{i=1}^{M} \mathcal{L}(x_i,y_i),
\end{equation}
where the loss for one record, $\mathcal{L}(x_i,y_i)$, is defined in \Cref{eq:loss_per_attribute}.

\subsection{Semi-supervised approach}
As mentioned in the introduction, the improvement phase of a PDI process often involves manual data cleaning via, e.g. user feedback or inspection by domain experts. Our basic approach does not require such manual cleaning, and the corresponding machine learning problem is unsupervised.

\begin{figure}
    \centering
    \includegraphics[width=0.8\textwidth]{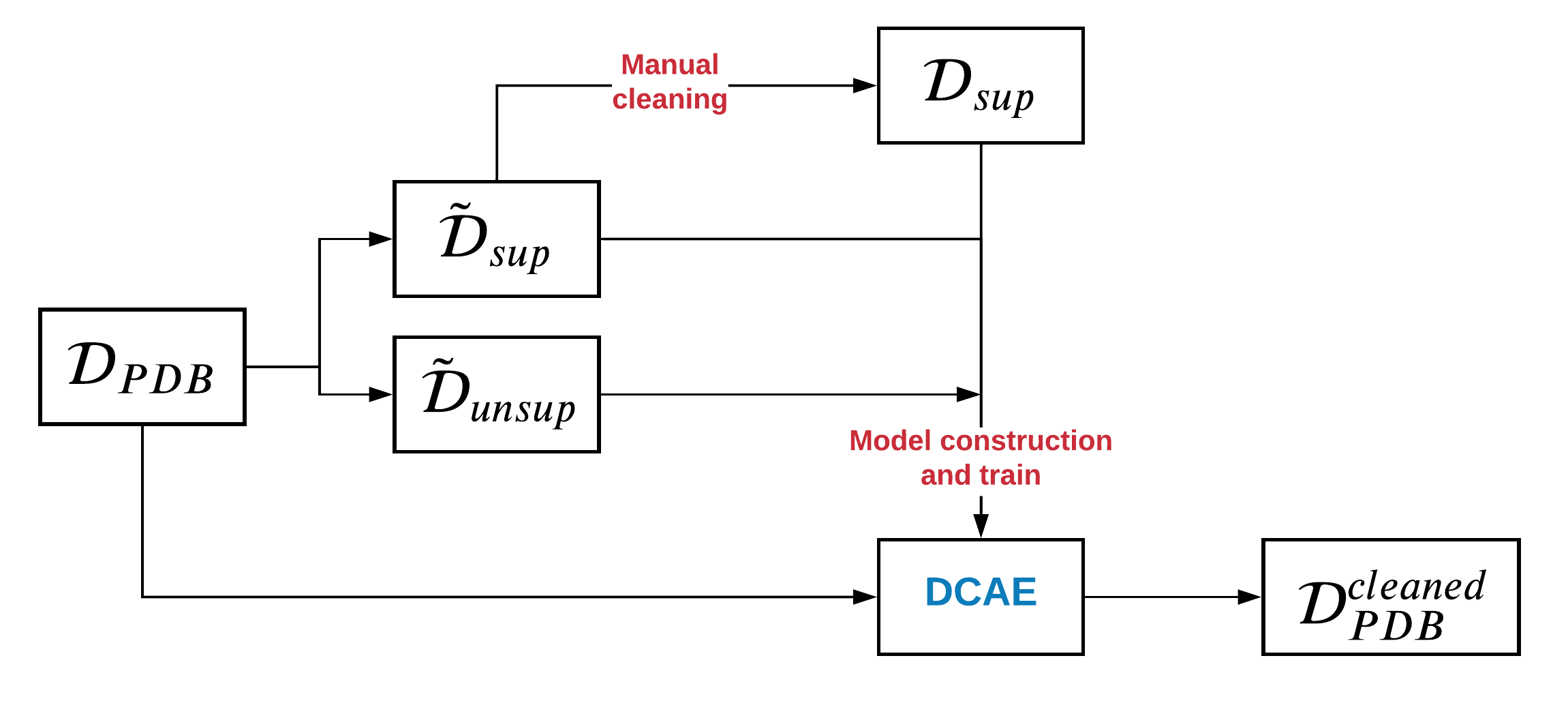}
    \caption{Semi-supervised training.}
    \label{fig:semi_sup}
\end{figure}

In addition to this unsupervised approach, we also investigate the performance of including a small fraction of manually cleaned records.
Besides learning the DCAE model to improve the data quality in an unsupervised setting, we thus also investigate the performance of the DCAE model in a semi-supervised setting. This means that the DCAE is trained on and applied to a PDB for which we know for a (small) subset what the outcome should be. In other words, given the probabilistic data $\mathcal{D}_{PDB}$, we partition it into a set $\tilde{\mathcal{D}}_{unsup}$ for which we do not know the ground truth, and a set $\tilde{\mathcal{D}}_{sup}$ for which we do know the ground truth. These ground truth labels are denoted with $\mathcal{D}_{sup}$. This is a semi-supervised setting: the DCAE is given $\tilde{\mathcal{D}}_{unsup}$ and $\tilde{\mathcal{D}}_{sup}$ as input and is trained to return $\tilde{\mathcal{D}}_{unsup}$ and $\mathcal{D}_{sup}$, respectively. This is depicted in \Cref{fig:semi_sup}.

\subsection{Extension to continuous attributes}\label{sec:categorization}
So far, both the PDB model in \Cref{sect.pdb_modeling} and the description of the input and output representation from \Cref{ssec:input_output} require the data to be categorical. In order to use this framework for continuous data we quantize the continuous attributes, resulting in discrete, categorical data. More precisely, for each continuous attribute $j\in\mathcal{A}$, the sample space is partitioned into $K_j$ bins, resulting in a histogram representation of the uncertainty.

More precisely, suppose attribute $j$ is taking values in the interval $[a,b]$. We perform binning based on $K_j+1$ thresholds $L_k$, $k=0, 1, \dots, K_j$ that satisfy
\begin{equation}
    a = L_0 < L_1 < \dots < L_{K_j} = b.
\end{equation}
In our experiments we work with 
$L_k = a+k\cdot\frac{b-a}{K_j},\ k=0,1,\ldots,K_j$ and we assign the value $B_k :=\frac{L_{k-1}+L_{k}}{2}$ to bin $k,\ k=1,\ldots,K_j$.

Now, if the value in record $i$ is represented as a random variable with cumulative distribution function (CDF) $F_{ij}(x)$, the resulting categorical probability distribution is
\begin{equation}
    p_{ij}(k)= F_{ij}(L_{k}) - F_{ij}(L_{k-1})  , \quad k=1,2,\ldots,K_j.
\end{equation}
In this setting, we refer to $K_j$ as the \emph{sampling density} for attribute $j$.

%% file: methodology.tex
\section{Methodology} \label{sec:methodology}

\begin{figure} 
    \centering
    \includegraphics[width=0.8\textwidth]{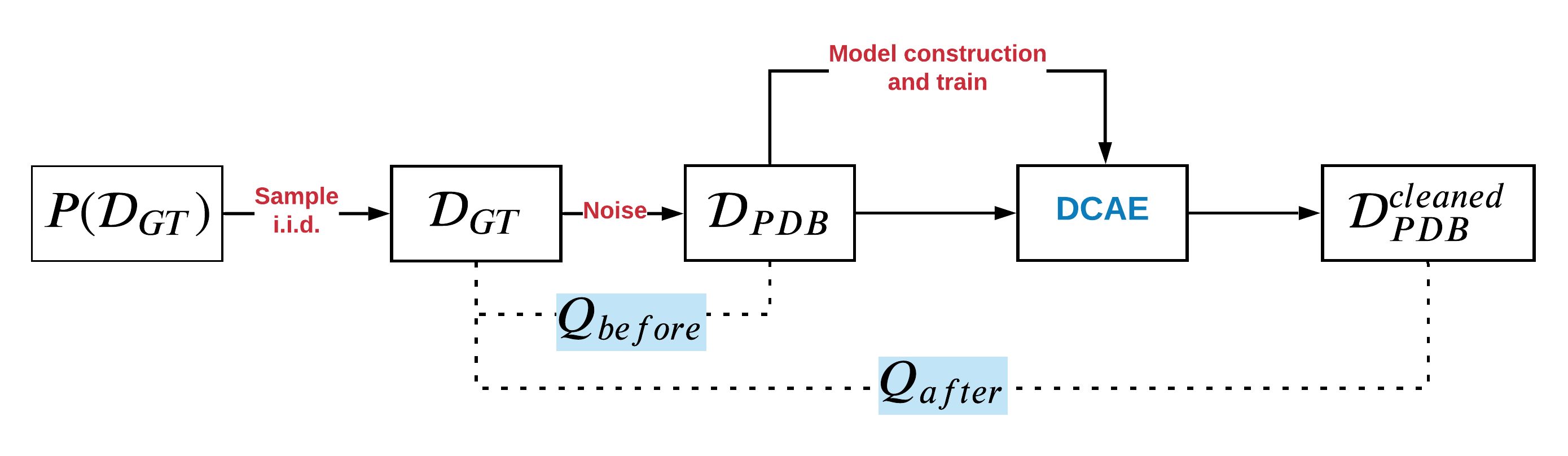}
    \caption{Evaluation process for the unsupervised setting.}
    \label{fig:eval_unsup}
\end{figure}

\begin{figure} 
    \centering
    \includegraphics[width=0.8\textwidth]{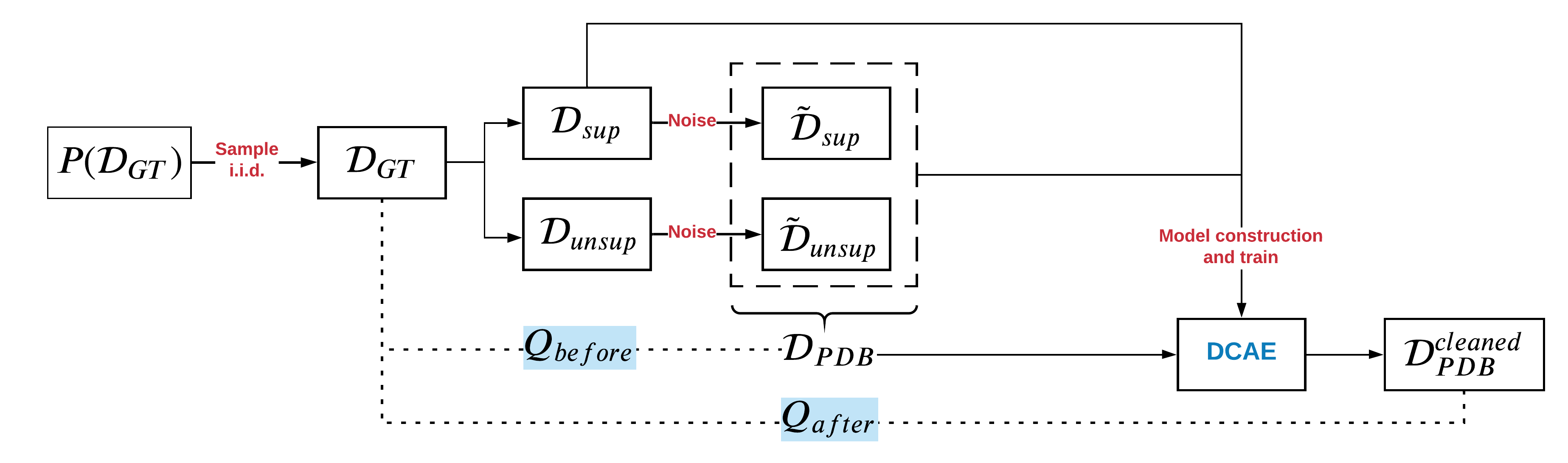}
    \caption{Evaluation process for the semi-supervised setting.}
    \label{fig:eval_semi}
\end{figure}

In this section we describe a methodological framework for evaluating the performance of DCAE. This framework relies on a comparison with a ground truth. Therefore, we necessarily work with synthetic data. In Section~\ref{sec:experiments} we support the experiments in this framework with experiments on real-world data.

\subsection{Overview of performance evaluation framework}\label{ssec:evaluation}
To get insight into the behaviour of our proposed solution and the impact of various design choices, we perform experiments in a well-controlled setting with synthetic data. We use a different performance valuation framework for the unsupervised and semi-supervised setting. An overview of these frameworks is given in \Cref{fig:eval_unsup} and \Cref{fig:eval_semi}. The details are provided below.

The elements of these frameworks are as follows:
\begin{enumerate}
    \item We start from a synthetic database in which there is no uncertainty and no errors. We refer to this data as our ground truth $\mathcal{D}_{\text{GT}}$ data, which can be seen as a data set that is sampled i.i.d. from an underlying ground truth data generating distribution $P(\mathcal{D}_{\text{GT}})$. We provide more details in \Cref{ssec:groundtruth}.
    \item
    
    We split $\mathcal{D}_{\text{GT}}$ into a labeled set and an unlabeled set, denoted as $\mathcal{D}_{\text{sup}}$ and $\mathcal{D}_{\text{unsup}}$, respectively. For unsupervised learning, $\mathcal{D}_{\text{unsup}} = \mathcal{D}_{\text{GT}}$ and $\mathcal{D}_{\text{sup}} = \emptyset $.
    \item Both $\mathcal{D}_{\text{sup}}$ and $\mathcal{D}_{\text{unsup}}$ are corrupted through the same noise process that yields $\Tilde{\mathcal{D}}_{\text{sup}}$ and $\Tilde{\mathcal{D}}_{\text{unsup}}$, respectively. This noise represents the uncertainty in the PDB. Together, they form $\mathcal{D}_{\text{PDB}}$, a corrupted version of the ground truth, similar to real-world probabilistic databases. In \Cref{ssec:addingnoise} we provide more details on this noise and how it is added.   
    \item The DCAE model is trained using $\Tilde{\mathcal{D}}_{\text{unsup}}$ and $\Tilde{\mathcal{D}}_{\text{sup}}$. Note, that $\mathcal{D}_{\text{sup}}$ corresponds to the manually cleaned version of $\Tilde{\mathcal{D}}_{\text{sup}}$. Therefore, we additionally use $\mathcal{D}_{\text{sup}}$ in the supervised setting.
    Note that in the unsupervised setting we train only on data \emph{with} noise, because this is what is typically available in practice. In the semi-supervised setting we include ground truth knowledge $\mathcal{D}_{\text{sup}}$ for part of the database.
    \item The performance of the DCAE model is tested by applying the trained DCAE to $\mathcal{D}_{\text{PDB}}$ so that we obtain $\mathcal{D}^{\text{cleaned}}_{\text{PDB}}$. We evaluate the improvement in data quality by comparing two values; on the one hand we measure the difference between $\mathcal{D}_{\text{GT}}$ and the original PDB data $\mathcal{D}_{\text{PDB}}$. We refer to this values as $\mathcal{Q}_{\text{before}}$. On the other hand we measure the difference between $\mathcal{D}_{\text{GT}}$ and the cleaned PDB data $\mathcal{D}^{\text{cleaned}}_{\text{PDB}}$. We refer to this value as $\mathcal{Q}_{\text{after}}$. By comparing $\mathcal{Q}_{\text{before}}$ and $\mathcal{Q}_{\text{after}}$, we can measure the performance of the DCAE. For more details 
        we refer to \Cref{ssec:perfmeasure}.
    
\end{enumerate}
Note that for evaluation, we do not use a test-train split. The method is intended to learn from a given real-world data set in order to clean 100\% of this same data set, instead of learning a model for evaluation on \emph{unseen data}.
Note also that a DCAE can easily and should be retrained on different datasets,
as the underlying distributions and the size of the AE's (corresponding to the number of columns in the database) are probably different. 
Furthermore, we want to emphasize that ground truth knowledge (i.e., labelled data $\mathcal{D}_{\text{sup}}$) is only used for training in the semi-supervised setting. 
In the unsupervised setting, the ground truth database $\mathcal{D}_{\text{GT}}$, and its partitions $\mathcal{D}_{\text{unsup}}$ and $\mathcal{D}_{\text{sup}}$, are not used in any way to train the DCAE, but only to measure the DCAE performance and for synthetic generation of data of lesser quality. Using the loss functions of \Cref{ssec:loss},
we can choose hyperparameters that maximize the DCAE's performance regardless of the underlying databases or distributions.

\subsection{Generating $\mathcal{D}_{\text{GT}}$} \label{ssec:groundtruth}

We use Bayesian networks (BN)~\cite[Chapter~8.1]{bishop2006pattern} to represent the data generating distribution $P(\mathcal{D}_{\text{GT}})$ which is used to generate synthetic data for the experiments. Such a network represents a set of variables (PDB attributes in our situation) and their conditional dependencies by means of a directed acyclic graph.
The variables in the BN can be categorical as well as numerical. While categorical variables and their realizations can be directly used in our PDB framework (\Cref{sect.pdb_modeling}), we first need to quantize our numerical variables to categorical variables (\Cref{sec:categorization}).

For most of our experiments, we use the BN $A\longrightarrow B\longrightarrow C$, but in some cases, we extend this chain to more than three variables. In all cases, the BN's that are used have the following properties:
\begin{itemize}
    \item The first variable, $A$, is a truncated and quantized standard normal distributed random variable on the interval $[-2,2]$. 
        \item Variable $B$ is a truncated and quantized Gamma distributed random variable. Conditioned on $A=a$, we use the Gamma distribution with parameters $\frac{30a}{K_C}+1$ and $1$ for shape and scale, respectively. We truncate to the interval $[4, 5+\frac{30a}{K_C}]$. We can summarize this relationship in an intuitive way: if the measured value of A is high, B is likely to be high. However, if the measured value of A is low, B is also likely to be low.
            \item Variable $C$ conditioned on $B$ has the same distribution as variable $B$ conditioned on $A$.

\end{itemize}

\begin{table}[t]
\centering\footnotesize
    \begin{subtable}[t]{0.3\textwidth}
        
        \centering
        \begin{tabular}[t]{llll}
                 &            &            &            \\
                 & \textbf{A} & \textbf{B} & \textbf{C} \\ \hline
0	& 0	& 0	& 2 \\
1	& 0	& 1	& 1 \\
2	& 2	& 3	& 3 \\
3	& 2	& 2	& 2 \\

            ...  & ...        & ...        & ...        \\
9999	& 1	& 1	& 1
 
        \end{tabular}
        \caption{Example database sampled from a Bayesian Network.         }
        \label{tab:database}
    \end{subtable}
    \hfill
    \begin{subtable}[t]{0.6\textwidth}
        
        \centering
        \begin{tabular}[t]{lllllllllllll}
            \textbf{Attribute} & \multicolumn{4}{l}{\textbf{A}} & \multicolumn{4}{l}{\textbf{B}}                    & \multicolumn{4}{l}{\textbf{C}}                    \\
            \textbf{Category}    & \textbf{0}     & \textbf{1}  & \textbf{2} & \textbf{3}  & \textbf{0} & \textbf{1} & \textbf{2} & \textbf{3} & \textbf{0} & \textbf{1} & \textbf{2} & \textbf{3} \\ \hline
0	& 1	& 0	& 0	& 0	& 1	& 0	& 0	& 0	& 0	& 0	& 1	& 0 \\
1	& 1	& 0	& 0	& 0	& 0	& 1	& 0	& 0	& 0	& 1	& 0	& 0 \\
2	& 0	& 0	& 1	& 0	& 0	& 0	& 0	& 1	& 0	& 0	& 0	& 1 \\
3	& 0	& 0	& 1	& 0	& 0	& 0	& 1	& 0	& 0	& 0	& 1	& 0 \\
            ...  & ... & ... & ... & ... & ... & ... & ... & ... & ... & ... & ... & ... \\            
            9999	& 0	& 1	& 0	& 0	& 0	& 1	& 0	& 0	& 0	& 1	& 0	& 0

        \end{tabular}
        \caption{The certain data of \Cref{tab:database} transformed to a probabilistic representation.}
        \label{tab:hard_evidence}
    \end{subtable}
     \caption{From sampled data to PDB.
     }
     \label{tab:evidences}
\end{table}

\begin{table}\footnotesize
    
    \centering
    
    \begin{tabulary}{\textwidth}{lllllllllllll}
        \textbf{Attribute} & \multicolumn{4}{l}{\textbf{A}} & \multicolumn{4}{l}{\textbf{B}}                    & \multicolumn{4}{l}{\textbf{C}}                    \\
        \textbf{Category}    & \textbf{0}     & \textbf{1} & \textbf{2} & \textbf{3}   & \textbf{0} & \textbf{1} & \textbf{2} & \textbf{3} & \textbf{0} & \textbf{1} & \textbf{2} & \textbf{3} \\ \hline
0 & 0.446 & 0 & 0.188 & 0.366 & 0.477 & 0.379 & 0.144 & 0 & 0 & 1 & 0 & 0 \\
1 & 1 & 0 & 0 & 0 & 0.16 & 0.531 & 0.002 & 0.307 & 0.089 & 0.573 & 0.001 & 0.337 \\
2 & 0 & 0 & 0.661 & 0.339 & 0.106 & 0.37 & 0.154 & 0.37 & 0.25 & 0.25 & 0.25 & 0.25 \\
3 & 0 & 0 & 0.597 & 0.403 & 0.213 & 0 & 0.489 & 0.297 & 0.352 & 0.286 & 0.362 & 0 \\
...  & ... & ... & ... & ... & ... & ... & ... & ... & ... & ... & ... & ... \\
9999 & 0.25	& 0.25	& 0.25	& 0.25	& 0.128	& 0.626	& 0.246	& 0	& 0	& 0.845	& 0.155	& 0
    \end{tabulary}
    
    \centering\caption{\centering The data from \Cref{tab:hard_evidence} with Gaussian noise added, and some missing entries by making all their probabilities equal.}
    \label{tab:noisy_data}
\end{table}

When running an experiment with more than three BN variables, we append a new variable $D$ to the bottom of the network, such as D, where $P(D|C) = P(C|B)$. We repeat this process for the next variables, i.e.: $P(E|D) = P(C|B)$, etc. 
Samples are taken from the joint probability distribution represented by this BN to generate $\mathcal{D}_\textit{GT}$. An example is given in \Cref{tab:database}. The input representation for the AE is generated by transforming this to a `one-hot encoding', because certain data is a special case of probabilistic data, where one value takes all the probability mass. See \Cref{tab:hard_evidence} for an example. 

\subsection{Corrupting our data with noise and errors} \label{ssec:addingnoise}

We model the data $\mathcal{D}_{\text{PDB}}$ residing in a PDB as a noisy version of the underlying ground truth data $\mathcal{D}_{\text{GT}}$, where noisy means that uncertainty (noise) is added to the ground truth data representative for data quality problems or imperfections in the data integration process. 
We start with `clean' records $x_i^{\text{GT}}\in \mathcal{D}_{\text{GT}}$ in the form of \Cref{tab:hard_evidence} and corrupt it producing noisy data as illustrated in \Cref{tab:noisy_data}.
We investigate two types of noise:

\begin{description}
    \item[Gaussian noise]
    
    We add Gaussian noise to $\mathcal{D}_{\text{GT}}$ by drawing and adding $\epsilon\sim N(0,\sigma$) to each cell in $\mathcal{D}_{\text{GT}}$, where we set negative entries to 0 and entries above 1 to 1. The probabilities are then normalized to sum to $1$.

    \item[Missing entry] Missing entry noise represents the realistic situation where a record contains missing entries. In a PDB with categorical data, this means that for a certain record $i$ and attribute $j$, we make the $K_j$ elements from $p_{ij}$ equal to each other, that is $p_{ij} = (\frac{1}{K_j}, \frac{1}{K_j},\ldots,\frac{1}{K_j})$, so that $p_{ij}$ contains zero knowledge about what category was observed. 
    
\end{description}

Gaussian noise is used in most experiments, as we expect that to be more prevalent than missing entries. In the last experiments shown in \Cref{sec:experiments}, we try to clean missing entries instead of removing Gaussian noise. 

We want to emphasize that we add unstructured noise to the data, i.e., noise is added independently over cells. Therefore, we do not introduce false patterns in the ground truth data.
The ground truth pattern is still present, so that it can be learned, despite the present noise.

The data in \Cref{tab:noisy_data} is an example of the data with the aforementioned noise, which we clean with our DCAE. An example of a result of cleaning \Cref{tab:noisy_data} can be found in \Cref{tab:denoised_data}.

\subsection{Performance measure} \label{ssec:perfmeasure}

As introduced in Section~\ref{ssec:evaluation} and illustrated in Figures~\ref{fig:eval_unsup} and~\ref{fig:eval_semi} we evaluate performance using $\mathcal{Q}_{\text{before}}$ and $\mathcal{Q}_{\text{after}}$, which measure the difference between the probability distribution of the ground truth, the noisy data and the outputted data from the DCAE. We report results using two measures:
\begin{enumerate}
    \item For numerical and categorical data: The JSD applied to the entire data sets, taking the sum over all the attributes and records as described in Equations~\eqref{eq:loss_per_attribute}, \eqref{eq:jsd_as_two_kl} and \eqref{eq:loss_per_PDB}. 
    \item For \emph{numerical} data: The rescaled MSE of the expected value of the probability distribution. Let $v_{ij} = \sum_{k\in\mathcal{K}_j}k\cdot p_{ij}(k)$ and $\Tilde{v}_{ij} = \sum_{k\in\mathcal{K}_j}k\cdot q_{ij}(k)$.
    We then define the rescaled MSE as
    \begin{equation}
        \mathcal{L}_j(p_{ij},q_{ij}) =\Big(\frac{v_{ij}-\Tilde{v}_{ij}}{B_{K_j}-B_{1}}\Big)^2, 
    \end{equation}
    where $B_{K_j}$ and $B_1$ are the largest and smallest bin values, see Section \ref{sec:categorization}.
    We do this for all attributes in all records and aggregate as in Equations~\eqref{eq:loss_per_attribute} and~\eqref{eq:loss_per_PDB}.
\end{enumerate}
We report on data quality improvement between the old and newly updated probabilistic data $\mathcal{D}_{PDB}$ and $\mathcal{D}^{cleaned}_{PDB}$, respectively, using the following measure:
\begin{equation}\label{eq.noise_red}
\text{Quality improvement in }\% = 100 - \Big(\frac
{  \mathcal{Q}_{\text{after}}  }
{   \mathcal{Q}_{\text{before}}  } \cdot 100\Big), 
\end{equation}
with $\mathcal{Q}_{\text{before}}$ and $\mathcal{Q}_{\text{after}}$ given by JSD or MSE. The higher this value is, the better the network performed, with a maximum of 100\% (meaning that all the noise was removed). If this value is below 0, the network was unable to remove noise and added noise to the dataset instead.

In addition, for \emph{categorical} data we report accuracy and F1 scores as follows. Let $p_{ij}^{GT}$ denote the ground truth value of a cell. We reduce uncertain attributes in $\mathcal{D}_{PDB}$ and $\mathcal{D}^{cleaned}_{PDB}$ to a maximum likelihood estimate by taking $p^{\text{max}}_{ij}:=\argmax_{k\in\mathcal{K}_j}p_{ij}(k)$ and $q^{\text{max}}_{ij}:=\argmax_{k\in\mathcal{K}_j}q_{ij}(k)$, respectively. Next, we identify correct `flips' of the data in $\mathcal{D}^{cleaned}_{PDB}$ as the case that $p_{ij}^{\text{max}}\neq p_{ij}^{GT}$ and $q^{\text{max}}_{ij}\neq p^{\text{max}}_{ij}$ and denote this as a True Positive. Similarly, we have
    \begin{equation}
        \begin{array}{c|cc}
            & q^{\text{max}}_{ij}=p^{\text{max}}_{ij} & q^{\text{max}}_{ij}\neq p^{\text{max}}_{ij} \\
            \hline
            p_{ij}^{\text{max}}=p_{ij}^{GT} & \text{True Negative} & \text{False Positive} \\
            p_{ij}^{\text{max}}\neq p_{ij}^{GT} & \text{False Negative} & \text{True Positive,} \\
        \end{array}
    \end{equation}
providing a complete binary classification test for which we report accuracy and F1 scores.

\subsection{Experimental setup and hyperparameters}\label{hyperparameters}

\begin{table}\centering\footnotesize
    \begin{subtable}[T]{0.55\linewidth}         
        \centering
        
        \begin{tabulary}{\linewidth}{lL}
        \textbf{Hyperparameter} & \textbf{Value}            \\ \hline
        Epochs                  & 100                       \\
        Batch size                  & 32                       \\
        Optimizer               & Adam \cite{Kingma2015}                      \\
        Training method         & Either semi-supervised (100 epochs unsupervised followed by 100 epochs supervised) or fully unsupervised           \\
        Activation types        & Sin, cos, linear, ReLU, Swish \\
        Hidden layers           & 3                         \\
        Latent space dimensions & Equal to BN size ($N$)         \\
        Loss function           & JSD      \\
        Activity regularizer    & L2, $(\lambda=10^{-4})$   \\
        Input layer type        & Gaussian noise           \\
        $\sigma_{\text{Gaussian noise layer}}$    & $0.01\cdot(100/K_j)$                 \\
        \end{tabulary}
        
        \caption{DCAE hyperparameters.}
        \label{tab:hyper_dcae}
    \end{subtable}
    \hfill     \begin{subtable}[T]{0.35\linewidth}
        
        \centering
        
        \begin{tabulary}{\linewidth}{LL}
        \textbf{Database parameter}   & \textbf{Value}            \\ \hline
        BN size ($N$)                        & 3                                    \\
        Sampling density ($K_j$)                & 4 ("categorical") or 100 ("continuous")  \\
        Records & 10000 \\
        Gaussian noise              & Enabled                                 \\
        $\sigma_{\text{PDB}}$   & $0.02\cdot(100/K_j)$            \\
        Missing entry noise               & Disabled                                \\
        Amount of labeled data for semi-supervised learning       & 2\%                                    
        
        \end{tabulary}
        
        \caption{PDB hyperparameters.}
        \label{tab:hyper_pdb}
    \end{subtable}
     \caption{Default hyperparameters for experiments.}
     \label{tab:hyperparameters}
     
\end{table}
For the experiments, we used TensorFlow and Keras to train the DCAE model, pyAgrum \cite{hal-02911619} for modelling BN's, and Pandas for operating on databases. We used a batch size of 32, with training split into 100 mini-batches, as these values led to the best trade-off between speed and performance. Each row $x_i$ in a PDB is one data point for training, where it is converted to a 1D input tensor compatible with the DCAE. Its length is equal to the amount of columns $\sum_{j=1}^{N}K_j$ in the PDB. However, as in most cases, we use a PDB with 10000 records and a batch size of 32 for training, the actual inputs used during training are 2D tensors with dimensions $\left(32, \sum_{j=1}^{N}K_j\right)$, as the first dimension in Keras/TensorFlow is the batch dimension. For training we use the Adam optimizer \cite{Kingma2015} to perform gradient descent. 

In most experiments, except those where we explicitly vary the sampling density, we use $K_j=4$ and $K_j=100$ as 
sampling densities for categorical and continuous attributes, respectively. Note that each attribute $j$ adds $K_j$ neurons to the input and output layers. Hence, more attributes or higher sampling densities increase the size of the network, which has a negative impact on training time and cleaning performance. Therefore, there is a trade-off for continuous attributes between a close approximation of the value, which calls for a high $K_j$ and a DCAE that is still small enough to train quickly and clean well.

Unless stated otherwise, we use a variance of   $\sigma_{\text{PDB}}=0.02\cdot(100/K_j)$ for the Gaussian noise, which is a data corruption ratio that often leads to the true value not being recognizable anymore by the naked eye and keeps the amount of noise added to the database independent of the sampling density.

In all cases, we use the Jensen-Shannon divergence (JSD) loss function.
Exploratory experiments using other loss functions, such as the mean square error (MSE), Kullback-Leibler (KL) divergence, and the categorical cross-entropy (log loss) were not promising and are not included due to space constraints.

\subsubsection{Experimental setup of DCAE architectural and hyperparameter experiments}

The goal of this set of experiments was to make a design choice or to establish a best setting for model (hyper) parameters/architecture. These include e.g. activation functions, regularization methods, noise parameters for a denoising AE architecture, etc. Due to space constraints we don't explicitly document these experiments.  The resulting hyperparameter values that followed from these experiments were used in the remaining experiments of Section \ref{sec:experiments} and can be found in Table \ref{tab:hyper_dcae}.
Details of these experiments can be found in~\cite{r_r_mauritz_2021_4603587}.

\subsubsection{Experimental setup of database parameter experiments}

The goal of this set of experiments is to measure the cleaning behaviour of the DCAE under varying data quality and database parameters.
\begin{itemize}
\item \emph{Experiment 1: Changing $\sigma_{\text{PDB}}$}\\
We vary the amount of Gaussian noise $\sigma_{\text{PDB}}$ added to the source data between $0.01\cdot(100/K_j)$ and $0.2\cdot(100/K_j)$.
\item \emph{Experiment 2: Adding missing entry noise without Gaussian noise}\\
As mentioned in \Cref{ssec:addingnoise}, we also introduce and experiment with missing values. We vary the number of missing values between 0.1\% and 50\%, while not adding any Gaussian noise.
\item \emph{Experiment 3: Combining missing entry noise and Gaussian noise}\\
To also investigate the effect of combining the two types of noise, we conduct the previous experiment with the presence of Gaussian noise.
\item \emph{Experiment 4: Changing the number of records in the database}\\
To see how well the DCAE solution scales to larger datasets, we investigate the effect of varying the number of records in the database that we train and evaluate the DCAE on.
\item \emph{Experiment 5: Changing the sampling density $K_j$}\\
We investigate the effect of the sampling density $K_j$ for continuous variables and the number of possible values for a categorical variable by varying $K_j$ between 4 and 300.
\item \emph{Experiment 6: Changing the BN size $N$}\\
Finally, we experiment with the size of the database in terms of the number of attributes $N$. Note that the number of attributes is the same as the BN size. We vary $N$ between 2 and 30. We expect to see similar effects as in Experiment~13 because the main consequence of a larger $N$ is similar to the main consequence of a larger sampling density $K_j$: a larger input and output layer.
\item \emph{Experiment 7: Varying the amount of labelled data for semi-supervised training}\\
We vary the amount of labelled data for semi-supervised learning from 0\% to 100\%. Note that in other experiments in which we report on the performance of semi-supervised training, we use $2\%$ labelled data.
\end{itemize}

\subsubsection{Experimental setup of real-world data experiments}

The goal of these experiments is to determine whether the DCAE - with the hyperparameters we chose after looking at the synthetic data - can actually be used to remove noise from real-world data. In addition to cleaning this data, we perform experiments by adding additional noise and missing entries to these datasets.

For this, we use databases obtained from real-life scenarios
and add more noise to them. These databases might already contain noise, meaning that choosing to use them as a "ground truth" may not be completely accurate, and this might lead to skewed results. However, the results we find should still give an indication of whether these techniques can be used for real data.

\begin{itemize}
\item \emph{Experiment 8: Adding Gaussian noise to real-world data}\\
We vary the amount of Gaussian noise $\sigma_{\text{PDB}}$ added to real-world data 
between $0.01\cdot(100/K_j)$ and $0.2\cdot(100/K_j)$.
\item \emph{Experiment 9: Adding missing entry noise and Gaussian noise to real-world data}\\
For this experiment, we add both Gaussian noise and missing entries to real-world data. $\sigma_{\text{PDB}}$ is left at its default value, while we vary the number of missing values between 0.1\% and 50\%.
\item \emph{Experiment 10: Adding missing entry noise to real-world data}\\
We introduce missing values to real-world data, but leave out Gaussian noise. We then vary the number of missing values between 0.1\% and 50\%.
\end{itemize}

An overview of the hyperparameters used in our experiments can be found in \Cref{tab:hyperparameters}.
The source code and data used for this research, including the \emph{complete} experimental setup, are open-source and can be found at \cite{r_r_mauritz_2021_4603587}.

%% file: experiments.tex
\section{Experiments and Results}\label{sec:experiments}

\subsection{Example results for synthetic databases}

\begin{table}\footnotesize
    
    \centering
    
    \begin{tabulary}{\textwidth}{lllllllllllll}
        \textbf{Attribute} & \multicolumn{4}{l}{\textbf{A}} & \multicolumn{4}{l}{\textbf{B}}                    & \multicolumn{4}{l}{\textbf{C}}                    \\
        \textbf{Category}    & \textbf{0}     & \textbf{1} & \textbf{2} & \textbf{3}   & \textbf{0} & \textbf{1} & \textbf{2} & \textbf{3} & \textbf{0} & \textbf{1} & \textbf{2} & \textbf{3} \\ \hline
0 & 0.965	& 0.026	& 0.007	& 0.002	& 0.731	& 0.238	& 0.016	& 0.015	& 0	& 0.98	& 0.016	& 0.004\\
1 & 0.999	& 0.001	& 0	& 0	& 0.276	& 0.643	& 0.008	& 0.072	& 0.031	& 0.353	& 0.21	& 0.405\\
2 & 0.041	& 0.03	& 0.888	& 0.041	& 0.009	& 0.082	& 0.12	& 0.79	& 0.006	& 0.009	& 0.406	& 0.579\\
3 & 0.054	& 0.017	& 0.861	& 0.068	& 0.015	& 0.008	& 0.841	& 0.136	& 0.006	& 0.012	& 0.928	& 0.055\\
...  & ... & ... & ... & ... & ... & ... & ... & ... & ... & ... & ... & ... \\
9999 & 0.391	& 0.569	& 0.035	& 0.006	& 0.009	& 0.962	& 0.021	& 0.009	& 0.001	& 0.641	& 0.342	& 0.016
    \end{tabulary}
    
    \caption{The result of training and evaluating the DCAE on the data from \Cref{tab:noisy_data}.}
    \label{tab:denoised_data}
\end{table}

In \Cref{tab:denoised_data} we show the result of evaluating the DCAE (with the default hyperparameters as described in \Cref{hyperparameters}) on a PDB (Table \ref{tab:noisy_data}) with both Gaussian noise ($\sigma_{\text{PDB}}=0.02\cdot\frac{100}{K_j}$) and missing entry noise (with the probability of entries missing at 5\%).

When comparing this table and \Cref{tab:noisy_data} to \Cref{tab:hard_evidence}, we see that the DCAE manages to drastically improve the data quality of most rows (such as row 3, which now has almost the same values as that same row in $\mathcal{D}_{\text{GT}}$ (Table \ref{tab:hard_evidence}), but with slightly more uncertainty).

It appears that the DCAE can often recover the ground truth (leading to distributions where the original value has a probability of $\geq0.9$), but this is not always the case. Sometimes, we wrongly introduce uncertainty (such as for row 9999, attribute $C$), or the DCAE barely removes any noise (such as in row 0, attribute $C$, where the original value cannot be recovered due to the large amounts of noise added). If the DCAE could always perfectly remove noise and never make any errors, we would see performance scores of near 100\%, so these errors are expected.

To show some more interesting properties of the DCAE, we look at the missing entries that were introduced for attribute C in row 2, and attribute A in row 9999. In both cases, the DCAE is able to recover the original value to a certain degree, as the category with the highest probability after cleaning is the category seen in the ground truth. For row 2, attribute C, it looks like the DCAE might have simply learned to replace the missing value for attribute C with the underlying distribution $P(C|B=3)$. However, for row 9999, attribute A, the distribution looks nothing like the original truncated normal distribution used to generate A. The DCAE has correctly inferred from the low values of B and C that A must also be high.

In the remainder of this section, we demonstrate through numerical experiments the influence of various design options and choice of hyperparameters. Additional results are available in \cite{essay80505}, \cite{essay82344}, and in the source code \cite{r_r_mauritz_2021_4603587}.

\subsection{Database parameter modifications}\label{ssec:parameter_mod}

\subsubsection{Experiment 1: Changing $\sigma_{\text{PDB}\label{sssec:experiment_9}}$}

\begin{figure}
     \centering
     \begin{subfigure}[t]{0.49\textwidth}
        \centering
        \includesvg[width=\textwidth]{Images/new/results_fig_sigma_plot_1.svg}
  \caption{JSD reduction for this experiment}
\label{fig:noise1}
     \end{subfigure}
      \hfill\begin{subfigure}[t]{0.49\textwidth}
        \centering
        \includesvg[width=\textwidth]{Images/new/results_fig_sigma_plot_2.svg}
  \caption{Accuracy and F1 score for flips on categorical data, and MSE reduction for continuous data for this experiment}
\label{fig:noise2}
     \end{subfigure}
     \caption{Effect of changing $\sigma_{\text{PDB}}$ on performance (\Cref{sssec:experiment_9}).}
    \label{fig:noise}
    
\end{figure}

It can be seen from \Cref{fig:noise} that the DCAE removes noise quite well when the amount of noise added is not too high. Only when there are high amounts of noise does the performance drop substantially (above $\sigma_{\text{PDB}}=0.02\cdot\frac{100}{K_j}$). It can be seen from \Cref{tab:noisy_data} that the noise added at $\sigma_{\text{PDB}}=0.02\cdot\frac{100}{K_j}$ is already a quite substantial. The fact that performance for semi-supervised training on high sampling densities remains quite high when adding even larger amounts of noise is quite remarkable. The performance seems to not decrease much further when even more noise is added, as there is already such a large loss of information that adding more noise does not affect cleaning performance. Furthermore, it is important to keep in mind that a sampling density of 100 is quite high; performance at lower sampling densities will be much better, as can be seen from \Cref{fig:sampling_density}.

It is important to note that even when the JSD reduction is not very high (as shown in \Cref{fig:noise1}), the accuracy and F1 score of flips, and the MSE reduction can still be quite high (as shown in \Cref{fig:noise2}). A good example is the performance of unsupervised training on continuous data when the amount of noise is low: the JSD reduction never rises above 75\%, but the MSE reduction goes up to 90\%.

\subsubsection{Experiment 2: Adding missing entry noise without Gaussian noise}\label{sssec:experiment_11}

\begin{figure}
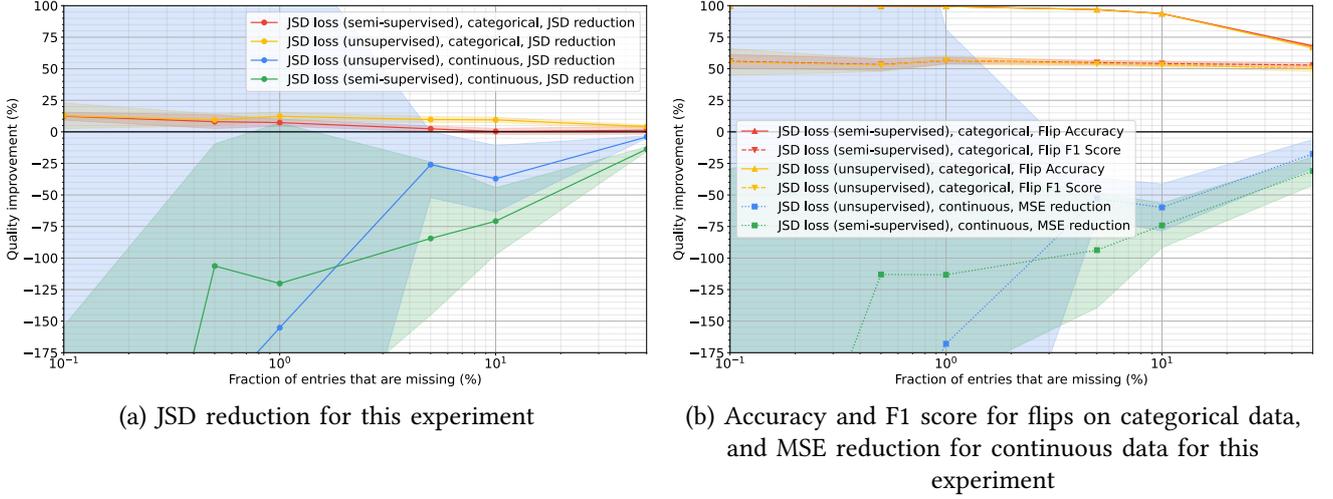

     \centering
     \begin{subfigure}[t]{0.49\textwidth}
        \centering
        \includesvg[width=\textwidth]{Images/new/results_fig_missing_entry_plot_1.svg}
  \caption{JSD reduction for this experiment}
\label{fig:missing_entry1}
     \end{subfigure}
      \hfill\begin{subfigure}[t]{0.49\textwidth}
        \centering
        \includesvg[width=\textwidth]{Images/new/results_fig_missing_entry_plot_2.svg}
  \caption{Accuracy and F1 score for flips on categorical data, and MSE reduction for continuous data for this experiment}
\label{fig:missing_entry2}      
     \end{subfigure}
    \caption{Effect of the likelihood of missing entry noise on performance, when $\sigma_{\text{PDB}}=0$ (\Cref{sssec:experiment_11}).}
    \label{fig:missing_entry}
    
\end{figure}

In \Cref{fig:missing_entry} we show what happens when we stop adding Gaussian noise to $\mathcal{D}_{\text{PDB}}$ and only add missing entries. The performance seems to be much lower when only looking at \Cref{fig:missing_entry1}.
The DCAE is clearly unable to compensate for missing entries in continuous data, adding more noise to the data in almost every case, with very unreliable performance (leading to a very large confidence interval). Performance for categorical data is not as bad, as the DCAE is always able to remove small amounts of noise.

{However, \Cref{fig:missing_entry2} shows that the performance for categorical data is better than can be seen from looking at just the JSD reduction alone. The accuracy is very high, as the DCAE has correctly learned to not 'flip' entries that are correct. The F1 score is an interesting metric as well here, as it does not take true negatives into account (the F1 score only uses values that were positive or predicted positive). A score of 50\% or higher is quite good, as it means the DCAE frequently picks the correct category out of the 4 possible categories to replace a missing entry with. It seems that the DCAE is an effective data imputation method.

One of our interpretations of this experiment was that the relatively low JSD reduction was caused by the fact that we were still using a Gaussian noise layer, even though there is no Gaussian noise in the data we were trying to clean. 
However, some further testing showed us the performance seen in \Cref{fig:missing_entry} always decreased when setting $\sigma_{\text{Gaussian noise layer}}$ to 0. We think that the low performance for continuous data can be explained by the fact that when $K_j$ is high, there are many possible bins the ground truth could occur in, and the probability of identifying the right bin from the other attributes goes down, especially due to the nature of the distributions we chose in \Cref{ssec:groundtruth}.

For the experiments below, we no longer show the accuracy, F1 score and MSE reduction for presentation reasons, as these plots did not show any important information that could not be inferred from the JSD reduction. These plots can still be found in the source code \cite{r_r_mauritz_2021_4603587}.

\subsubsection{Experiment 3: Combining missing entry noise and Gaussian noise}\label{sssec:experiment_10}

\begin{figure}
     \centering
     \begin{minipage}[t]{0.49\textwidth}
        \centering
        \includesvg[width=\textwidth]{Images/new/results_fig_missing_entry_combined_plot_1.svg}
    \caption{Effect of missing entries on performance, when $\sigma_{\text{PDB}}=0.02\cdot(100/K_j)$ (\Cref{sssec:experiment_10}).}
\label{fig:missing_entry_combined}
     \end{minipage}
     \hfill\begin{minipage}[t]{0.49\textwidth}
        \centering
        \includesvg[width=\textwidth]{Images/new/results_fig_rows_plot_1.svg}
  \caption{Effect of the amount of rows in the database on performance (\Cref{sssec:experiment_12}).}
\label{fig:rows}     
     \end{minipage}
\end{figure}

We can see in \Cref{fig:missing_entry_combined} that adding missing entry noise to a $\mathcal{D}_{\text{PDB}}$ that already has Gaussian noise ($\sigma_{\text{PDB}}=0.02\cdot(100/K_j)$) does not seem to affect performance much until 5\% of entries or more are missing. This is a very high amount of missing data. Even then, the performance is still quite good. Performance starts sharply decreasing for all cases when 10\% or more of the entries are missing. Even then, the performance in most cases is still relatively high. Performance is highest on continuous data; this is similar to the results seen in \Cref{fig:noise} for $\sigma_{\text{PDB}}=0.02\cdot(100/K_j)$ and only drops off when 10\% or more of the entries are missing, similar to the results seen in \Cref{fig:missing_entry}.

\subsubsection{Experiment 4: Changing the amount of records in the database}\label{sssec:experiment_12}

\Cref{fig:rows} shows us that the performance of the DCAE scales very well with the number of records when semi-supervised training is used. For continuous data, data cleaning performance reaches almost 100\% when the amount of records is high. This is probably because there is now 100 epochs of training on hundreds or even thousands of rows where the ground truth is available, instead of overfitting on only a few.

Unsupervised training does not seem to benefit from having more than 1000 rows; the performance stops rising at that point for both continuous and categorical data, although the DCAE seems to be able to clean more noise in continuous data. This can be explained by the fact that the DCAE is simply being taught to reproduce the input data, and 1000 rows is enough to exhaust most combinations of the three BN variables, with various amounts of noise added.

\subsubsection{Experiment 5: Changing the sampling density $K_j$}\label{sssec:experiment_13}

\begin{figure}
     \centering
     \begin{minipage}[t]{0.49\textwidth}
        \centering
        \includesvg[width=\textwidth]{Images/new/results_fig_sampling_density_plot_1.svg}
  \caption{Effect of sampling density ($K_j$) on performance (\Cref{sssec:experiment_13}).}
\label{fig:sampling_density}
     \end{minipage}
     \hfill
     \begin{minipage}[t]{0.49\textwidth}
        \centering
        \includesvg[width=\textwidth]{Images/new/results_fig_BN_size_plot_1.svg}
  \caption{Effect of BN size ($N$) on performance (\Cref{sssec:experiment_14}).}
\label{fig:bayesian_size}
     \end{minipage}

\end{figure}

As can been see from \Cref{fig:sampling_density}, performance seems to be highest at $K_j=50$. The fact that the performance decreases as the sampling density increases makes sense, as the dimensionality of the data increases and as a result, there are more trainable weights. The decrease in performance at very low sampling densities can also be explained by the fact that missing entries are introduced by the relatively large amounts of Gaussian noise per bin. Furthermore, at high sampling densities, it is easy for the network to introduce a very large amount of JSD "error" by outputting a distribution where the maximum probability is in the wrong bin/category, but only off by 1. This is because the JSD loss function penalizes just as much for choosing a value close to the true bin, as for choosing a value 200 bins away from the true value. We found a potential future work solution to this problem, that we discuss in \Cref{sec:conclusions}.

The fact that this configuration is able to reduce large amounts of noise at a sampling density of 100 is remarkable, as this has an input layer of 300~neurons ($3 \text{ attributes}\cdot100 \text{ bins}=300$). This is a decently large neural network, and being able to reach a good performance with only 100 epochs of training (which takes only a few minutes) is outstanding.

We also observed that at low sampling densities ($K_j=4$), it is much faster to train the network with a consumer-grade quad-core CPU than a consumer-grade GPU. This is probably caused by the overhead introduced by moving data to the GPU, which offers no speed-up due to the lack of large matrix multiplications for small neural networks. At higher sampling densities ($K_j\geq100$), this difference was not observed.

\subsubsection{Experiment 6: Changing the BN size ($N$)}\label{sssec:experiment_14}

Some of the trends that could be observed in \Cref{sssec:experiment_13} can also be seen in \Cref{fig:bayesian_size}. The performance of the DCAE seems to decrease as the amount of neurons in the DCAE increases, hence also when the number of attributes in $\mathcal{D}_{\text{PDB}}$ increases. The drop-off is much sharper at high sampling densities, as at $N=5$, there are already $5\cdot100=500$}neurons, while for low sampling densities at $N=5$, there are $5\cdot4=20$ neurons.
The drop off in performance here is not as sharp as the drop in performance seen in \Cref{fig:sampling_density}, for categorical data and for unsupervised training on continuous data. This is because the middle layer of the network scales with $N$, adding more neurons and trainable weights to the network. The increase in the network size is the reason why it is still able to learn to clean such large input tensors.
Performance for semi-supervised learning on continuous data decreases as expected for large values of $N$, as this leads to input tensors of sizes of 1000 or higher. However, unsupervised learning performance does not seem to be affected as much by this. %

\subsubsection{Experiment 7: Varying the amount of labeled data for semi-supervised training}\label{sssec:experiment_7}

\begin{figure}
        \centering
        \includesvg[width=0.5\textwidth]{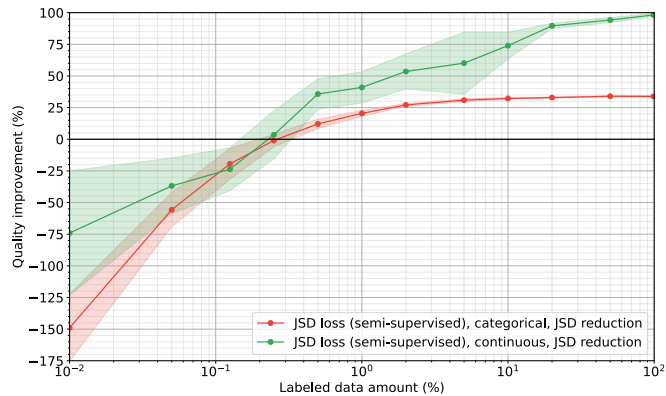}
  \caption{Effect of the amount of labeled data on semi-supervised training performance (\Cref{sssec:experiment_7}).}
\label{fig:label_amount}
\end{figure}

In \Cref{fig:label_amount}, we show the effect of changing the amount of labelled data.
Here, we expected 100\% labelled data to be the same as supervised learning, while 0\% labelled data should theoretically be the same as unsupervised learning. (The Keras/TensorFlow implementation we used did not allow us to set values of 0\% and 100\%, so we used 0.01\% and 99\% instead).

Performance at high amounts of labelled data seems to be very high, as would be expected from supervised learning. However, we witness diminishing returns for the amount of labelled data we provide: labelling more than 10\% of the data does not seem to affect the performance at all; at 5\%, it is not much worse.
Thus, it seems that this solution does not require a lot of labelled data for good performance.
Choosing 2\% for the other experiments appears not only to be a realistic setting, but also one that still achieves good cleaning performance.

Interestingly, performance at very low amounts of labelled data does not at all match the performance of unsupervised learning in other experiments. We suspect that this is caused by overfitting, as in the most extreme case, we perform unsupervised training on 99.9\% of the data for 100 epochs, followed by 100 epochs of supervised training on 0.01\% of the data. In a database of 10000 rows, this is merely one row, which could explain the drop in performance.

\subsection{Experiments on real-world data}\label{ssec:real_world_exps}
In this section we present results on real-world datasets.
\subsubsection{Processing real-world data}\label{sssec:processing_real_world}

The first dataset that we use contains data on surgeries conducted at a hospital, with information such as the type and duration of the surgery. The second dataset contains data from a questionnaire filled out by patients with chronic pain as well as the treatment they received based on their answers.
Before converting them to PDBs, these datasets were preprocessed by replacing numerical intervals like 20--29 and 30-39 by single values like 20 and 30, and by replacing entries that contain words such as "unknown" (or translations of it) by NULL values, and by adding "CATEGORICAL" to the header of each column with categorical data. This is to make sure that categorical data that is represented by numbers (such as foreign keys in a database) is not converted to a numerical distribution. 

We chose the amount of bins $K_j$ for a numerical attribute as a function of the number of unique values observed in the dataset. For $n$ unique values we let $K_j = \mathrm{min}\left\{n,\mathrm{max}\left(\mathrm{fd}\left(n\right), \mathrm{sturges}\left(n\right)\right)     \right\}$, where fd is the Freedman-Diaconis rule and sturges is Sturges' formula. This method was chosen because it leads to a more realistic output than when $K_j$ is constant. $\mathrm{max}\left(\mathrm{fd}\left(n\right), \mathrm{sturges}\left(n\right)\right)$ is the method numpy.histogram uses to calculate the amount of bins, while taking the minimum of that equation and and $n$ keeps discrete attributes from being turned into continuous ones.

\subsubsection{Example results for real-world data}\label{sssec:real_world_example_results}

In \Cref{tab:real_world_data_before} and \Cref{tab:real_world_data_after}, we show the results of unsupervised training on the chronic pain questionnaire data. We chose to show this dataset, as it contained mostly numerical data which is easily presented in this paper. Cleaning results on other data are available in the source code \cite{r_r_mauritz_2021_4603587}. As we do not know the ground truth for this data, we cannot perform semi-supervised training on it, and we cannot report the JSD reduction, accuracy and F1 scores, since the ``true values'' are not known.

\begin{table}[]\centering\footnotesize
\begin{adjustbox}{max width=0.75\textwidth}
\begin{tabulary}{\linewidth}{C|CCCCCCCCCCCCCCCCC}
& \rot{Treatment CATEGORICAL} & \rot{Fever CATEGORICAL} & \rot{Duration of pain} & \rot{Sick leave CATEGORICAL} & \rot{Earlier hospitalization CATEGORICAL} & \rot{Workoverload CATEGORICAL} & \rot{Familiy history CATEGORICAL} & \rot{Depression CATEGORICAL} & \rot{Extremely nervous} & \rot{Stress CATEGORICAL} & \rot{Relationship with colleagues} & \rot{Irrational thoughts risk lasting} & \rot{Irrational thoughts work} & \rot{Coping strategy} & \rot{Kinesiophobia physical exercise} & \rot{Kinesiophobia pain stop} & \rot{Age}   \\ \hline
0 & 1 & 1 & 10 & 0 & 1 &  & 0 & 0 & 0 & 0 &  & 9 & 10 & 0 & 7 & 7 & 70 \\
1 & 3 & 1 & 10 & 0 & 1 & 0 & 1 & 0 & 0 & 1 & 9 & 5 & 0 & 7 & 10 & 10 & 50 \\
2 & 1 & 1 & 10 & 0 & 1 &  & 0 & 0 & 6 & 1 & 4 & 8 & 8 & 2 & 7 & 7 & 50 \\
3 & 5 & 1 & 10 & 0 & 1 &  & 0 & 0 & 1 & 1 & 9 & 10 & 3 & 1 & 10 & 10 & 50 \\
4 & 1 & 1 & 10 & 0 & 1 &  & 0 & 0 & 0 & 1 & 8 & 7 & 10 & 7 & 7 & 9 & 50 \\
\end{tabulary}
\end{adjustbox}
\caption{The first rows and columns of the chronic pain questionnaire data, before cleaning.}
\label{tab:real_world_data_before}
\end{table}

\begin{table}[]\centering\footnotesize
\begin{adjustbox}{max width=0.75\textwidth}
\begin{tabulary}{\linewidth}{C|CCCCCCCCCCCCCCCCC}
& \rot{Treatment CATEGORICAL} & \rot{Fever CATEGORICAL} & \rot{Duration of pain} & \rot{Sick leave CATEGORICAL} & \rot{Earlier hospitalization CATEGORICAL} & \rot{Workoverload CATEGORICAL} & \rot{Familiy history CATEGORICAL} & \rot{Depression CATEGORICAL} & \rot{Extremely nervous} & \rot{Stress CATEGORICAL} & \rot{Relationship with colleagues} & \rot{Irrational thoughts risk lasting} & \rot{Irrational thoughts work} & \rot{Coping strategy} & \rot{Kinesiophobia physical exercise} & \rot{Kinesiophobia pain stop} & \rot{Age}   \\ \hline
0 & 1 & 1 & 10 & 0 & 1 & 0 & 0 & 0 & 0 & 0 & 1.68 & 9 & 10 & 0 & 7 & 7 & 70 \\
1 & 3 & 1 & 10 & 0 & 1 & 0 & 1 & 0 & 0.01 & 1 & 9 & 5 & 0.01 & 7 & 10 & 10 & 49.99 \\
2 & 1 & 1 & 10 & 0 & 1 & 0 & 0 & 0 & 6 & 1 & 4 & 8 & 7.99 & 2 & 7 & 7 & 50 \\
3 & 5 & 1 & 10 & 0 & 1 & 1 & 0 & 0 & 1 & 1 & 8.98 & 10 & 3 & 1 & 10 & 10 & 50 \\
4 & 1 & 1 & 10 & 0 & 1 & 0 & 0 & 0 & 0 & 1 & 8 & 7 & 10 & 7 & 7 & 9 & 50 \\
\end{tabulary}
\end{adjustbox}
\caption{The chronic pain questionnaire data, after cleaning using unsupervised learning.}
\label{tab:real_world_data_after}
\end{table}

We can see that the cleaning process does not change most existing values, as it has learned to reproduce those as accurately as possible. Any changes in existing values here can be explained by the centers of the bins are not being placed exactly on integer values. If it is important that values remain integers, the output data will require some manual postprocessing (rounding, for example).

Interestingly, the DCAE has produced believable values for all the missing entries. For example, the person in row 0 is probably a pensioner due to their age; they would not have a good relationship with co-workers (as they would no longer have any). The person described in row 3 reports feeling stress, being extremely nervous, and having a very good relationship with their coworkers (a 9/10). It seems likely that they would report a "work overload". In \Cref{par:experiment_realworld_2}, we see that these values are likely to be accurate.

\subsubsection{Adding noise to real-world data}\label{sssec:real_world_experiment_plots}

For the next experiments, we add more noise to the surgical case durations dataset, in the same way we would add noise to synthetic data as explained in~\Cref{ssec:addingnoise}. The results for the chronic pain questionnaire dataset (which is much smaller) are quite similar, and can be found in the source code \cite{r_r_mauritz_2021_4603587}.
The results from this should be interpreted accordingly; we do not know the actual ground truth of these datasets, as there is already noise in them. Note, that we do not add Gaussian noise to cells that already have a missing entry. We only show the MSE reduction, accuracy, and F1 score of flips for Experiment 10, as they do not affect our conclusions for the other experiments. However, more figures can be found in the source code \cite{r_r_mauritz_2021_4603587}.

\paragraph{Experiment 8: Adding Gaussian noise to real-world data}\label{par:experiment_realworld_1}

\begin{figure}
     \centering
     \begin{minipage}[t]{0.49\textwidth}
        \centering
        \includesvg[width=\textwidth]{Images/new/results_REALWORLD_sigma_surgical_case_durations__1.svg}
  \caption{Effect of changing $\sigma_{\text{PDB}}$ on performance for real-world data (\Cref{par:experiment_realworld_1}).}
\label{fig:real_world_sigma}
     \end{minipage}
\hfill\begin{minipage}[t]{0.49\textwidth}
        \centering
 \includesvg[width=\textwidth]{Images/new/results_REALWORLD_missing_entry_combined_surgical_case_durations__1.svg}
    \caption{Effect of missing entries on real-world data performance, when $\sigma_{\text{PDB}}=0.02\cdot\frac{100}{K_j}$ (\Cref{par:experiment_realworld_3}).}
\label{fig:real_world_missing_entry_combined}
     \end{minipage}     
\end{figure}

\Cref{fig:real_world_sigma} shows that the DCAE is quite good at removing large amounts of Gaussian noise when semi-supervised training is used. However, this is not the case for unsupervised training.

\paragraph{Experiment 9: Adding missing entry noise and Gaussian noise to real-world data}\label{par:experiment_realworld_3}

\Cref{fig:real_world_missing_entry_combined} shows that the performance for both methods of training is quite similar to the performance seen in Experiment 8, for $\sigma_{\text{PDB}}=0.02\cdot\frac{100}{K_j}$.

\paragraph{Experiment 10: Adding missing entry noise to real-world data}\label{par:experiment_realworld_2}

\begin{figure}
     \centering
     \begin{subfigure}[t]{0.49\textwidth}
        \centering
        \includesvg[width=\textwidth]{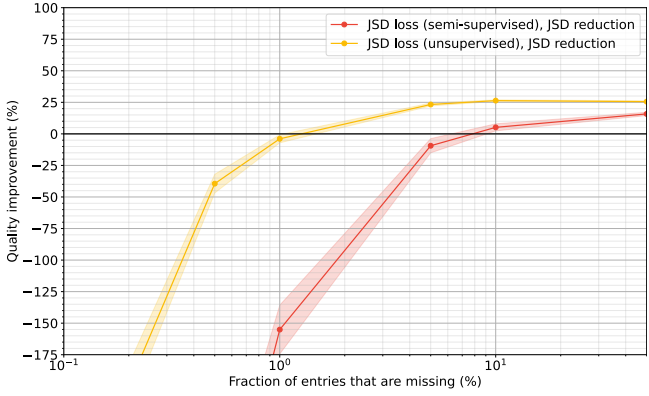}
  \caption{JSD reduction for this experiment}
\label{fig:real_world_missing_entry1}
     \end{subfigure}
\hfill\begin{subfigure}[t]{0.49\textwidth}
        \centering
        \includesvg[width=\textwidth]{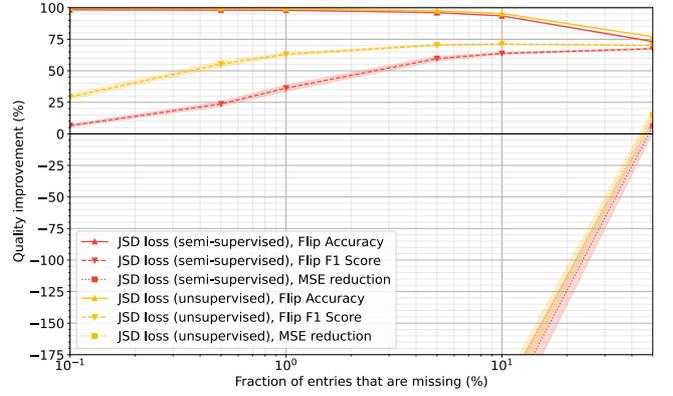}
  \caption{Accuracy and F1 score for flips, and MSE reduction for continuous data}
\label{fig:real_world_missing_entry2}
     \end{subfigure}     
    \caption{Effect of missing entries on real-world data performance, when $\sigma_{\text{PDB}}=0$ (\Cref{par:experiment_realworld_2}).}
    \label{fig:real_world_missing_entry}
\end{figure}

\Cref{fig:real_world_missing_entry} shows that the DCAE is not able to remove much noise (if measured using the JSD) when not many entries are missing, but this performance seems to improve as more entries are missing. For higher amounts of missing entries, the F1 score for flipped entries is above 50\% while the accuracy is above 95\%, which is quite significant. Interestingly, the performance achieved when using unsupervised training is much higher than the performance achieved using semi-supervised training. Also, the MSE reduction for continuous variables is practically never above 0: it seems that we are not good at reproducing numerical missing entries (which matches our findings in \Cref{sssec:experiment_11}).

The performance we see in this experiment is promising for the results seen in \Cref{sssec:real_world_example_results}. When about 7 percent of entries are missing (which is the case for the data used in \Cref{sssec:real_world_example_results}), the accuracy of flipped entries is about 95\%, and the F1 score is about 70\%. While the plots shown here are for a different dataset than the one shown in \Cref{sssec:real_world_example_results}, experiments for that dataset (for which plots can be found in the source code \cite{r_r_mauritz_2021_4603587}) show similar results, but with an F1 score closer to 65\%.

When looking at the last three experiments, we can deduce why the semi-supervised and unsupervised training methods lead to different performance levels on the different types of noise we introduce.

During unsupervised training, the DCAE learns to reproduce the input data as accurately as possible (while ignoring missing entries, as described in \Cref{hyperparameters}). This explains the high performance of this method on missing entries, as the DCAE should have learned to reproduce large amounts of input data, and be able to infer what the missing entry would have been. However, this strategy of accurately reproducing input data is not good for removing Gaussian noise.

During semi-supervised training, unsupervised training occurs for 100 epochs, after which supervised training occurs for 100 epochs on a small subset of the data. This might lead to some overfitting, but it will lead to the network becoming very good at getting rid of errors similar to those seen in the subset of labelled data. However, this will reduce its ability to accurately reproduce the input data, which may explain why semi-supervised learning does not perform as well for reducing missing entry noise, while leading to high performance for Gaussian noise.

%% file: conclusions.tex
\section{Conclusions}
\label{sec:conclusions}

In this paper, we propose an autoencoder-based data cleaning approach, referred to as DCAE, capable of near-automatic data quality improvement. The intuition behind the approach is that an autoencoder can learn structure and dependencies hidden in data, which can be used to identify and correct doubtful values. The approach uses a probabilistic data representation to express weak and strong doubts. It can be used for cleaning both ordinary `crisp' data sets as well as probabilistic data resulting from a probabilistic data integration process.
We also introduce a Bayesian Network-based approach for generating synthetic test data with embedded dependencies to evaluate our cleaning approach.

In our experiments, we varied the level of noise and missing values, number of attributes, the number of records, sampling density for continuous attributes, and several other hyper-parameters. We also experimented with several alternatives for the autoencoder architecture. Results show that fully automatic cleaning in an unsupervised manner is possible, but that a semi-supervised setting with a mix of unlabeled data and only little reference data (2\% in our experiments) produced significantly better results: for data sets with up to 4~attributes, 25\% removal of artificially introduced noise in categorical attributes and 40\%-65\% in continuous attributes. The performance increases further when more records are added to the database. Apparently, an autoencoder is able to learn structure and dependencies hidden in data sets, and we can exploit this for the purpose of data cleaning. The DCAE managed to restore missing entries in categorical data reasonably well, but was unable to do so for continuous data. When Gaussian noise is present and up to 10\% of entries are missing, performance is good for continuous and categorical data. The performance of the DCAE seems to be even better on real-world data, as the nature of the synthetic data we use prevents easily inferring missing attributes from other attributes. Best results were obtained with an architecture with a Gaussian noise layer, limited (up to 5) hidden layers, JSD loss function, a combination of sin, cos, linear, ReLU and Swish activation functions, and L2 activity regularization with $\lambda=10^{-4}$.

For future work, more experimentation is desired. First, it would be interesting to experiment with more complex dependencies by sampling data from Bayesian Networks $P(\mathcal{D}_{\text{GT}})$ with different non-linear structures. Second, it seems logical that longer training (we trained for 100 epochs) and more effective labelling approaches for semi-supervised learning (such as scaling the number of epochs for supervised training with the amount of labelled data), will improve performance. However, such expectations need to be validated before they can be used as recommendations. Third, the performance of the approach seems to degrade when the input and output layers grow beyond about 500~neurons. Architectural changes may counteract this disadvantageous behaviour.

Fourth, as was mentioned in Section \ref{sect.pdb_modeling}, our DCAE approach assumes that the categorical data is nominal. This, however, is problematic for the quantized continous data, as the original continuous data often \emph{does} contain an ordering. As a consequence, our DCAE approach now falsely evaluates the loss/performance measure for this quantized continuous data in case of using the JSD loss metric/performance measure. To see this, consider the example where an age-attribute is quantized into 0-25, 25-50, 50-75 and 75-100. Given that the ground truth age was e.g. 70, our DCAE approach equally penalizes the prediction 0-25 and 25-50, whereas the latter clearly is a better prediction. To better train the DCAE and evaluate the DCAE performance for continuous data, one could use a different loss function and performance measure than the JSD, such as the Wasserstein distance.

Preliminary results that can be found in the source code \cite{r_r_mauritz_2021_4603587} suggest that the Wasserstein distance is a vastly superior loss function for semi-supervised learning on numerical data, when compared to the JSD loss function. This leads to both a larger decrease in the Wasserstein distance itself, and a significant decrease in the MSE of the numerical entries after cleaning. This solution seems to scale much better with higher sampling densities, Bayesian network sizes, and large values of $\sigma_{\text{PDB}}$, but more investigation is needed to find out if it is truly viable for data cleaning.

Furthermore, the results seen in \Cref{sssec:real_world_experiment_plots} suggest that unsupervised training is good for removing missing entries, while semi-supervised training is good for removing Gaussian noise. Perhaps it is possible to achieve higher performance on datasets with both types of noise by applying both methods: first, removing missing entries by using unsupervised training on a DCAE and evaluating it on the data, then removing Gaussian noise in this cleaned data by using semi-supervised training on a new DCAE.

Finally, the largest challenge when applying the DCAE solution on real-world databases is that it is impossible to know whether the data cleaning was a success, as there is no ground truth data $\mathcal{D}_{\text{GT}}$ to compare to. Whether the DCAE successfully learned to improve data quality is not apparent from its output. A possible solution would be to incorporate techniques that also output confidence intervals, such as those used in Gaussian Process classifiers \cite{rasmussen2006gaussian}. Then, the user can see whether the DCAE was able to learn the underlying structures of the data. When the confidence intervals of the output are small, the output is likely to be correct, while when confidence intervals are large, the output is more likely to be wrong (as the DCAE is "unsure" about the underlying distribution).